%% file: main.tex
\documentclass[letterpaper,twocolumn,10pt]{article}
\usepackage{usenix}

\usepackage{tikz}
\usepackage{amsmath}
\usepackage{multirow}
\usepackage{color}
\usepackage{xcolor} 
\usepackage{url}
\usepackage{float}
\usepackage{xspace}
\usepackage{makecell}
\usepackage{makecell}
\usepackage{xcolor}
\usepackage{tcolorbox}
\usepackage{booktabs}
\usepackage{circledsteps}
\newcommand{\system}{\textsc{Priva-See}\xspace}
\usepackage{xurl}
\usepackage{hyperref}
\usepackage{cuted}
\usepackage{amssymb}

\begin{document}

\date{}

\title{\Large \bf Privacy Leakage Through AI-mediated Analysis of Smartphone Data}

\author{
{\rm Sarah Radway}\\
Harvard University
\and
{\rm Zoe Robert}\\
Unaffiliated
\and
{\rm Matthew Soto}\\
Carnegie Mellon University
\and
{\rm Julianna Cimillo}\\
Unaffiliated
\and 
{\rm Sebastian Diaz}\\
Harvard University
\and
{\rm Meg Marco}\\
Harvard University
\and
{\rm James Mickens}\\
Harvard University
} 

\maketitle

\begin{abstract}
Over the past thirty years,
the online advertising industry
built a large-scale data collection ecosystem,
with the goal of tracking a user's online activity 
to infer their demographics and interests. 
Traditionally,
the ecosystem relied upon the collation
and analysis of highly-structured text data
like user IP addresses, GPS coordinates, e-commerce purchase histories,
and visited URLs.
However, recent ML models can parse not only structured text, 
but also multimedia files and unstructured text inputs---meaning
a user's photos, videos, inboxes, and calendars 
are now ripe for automated analysis.
The privacy risks
are particularly acute in the context of smartphone apps.
A user's phone already acts as
a natural collation point for sensitive user information,
but users may not understand that
permitting an app to, for example,
access a user's photo
does not just give the app access to
the bytes in the photo:
the app also receives access to
\textit{inferences about the user}
that are enabled by the photo.

To explore these privacy risks,
we built \system,
an LLM-based inference system
for app-collected user data;
\system reflects our best understanding
of how real-life adtech companies
would leverage machine learning
to build user profiles.
Through an IRB-approved user study,
465 participants deployed \system on their phones;
\system made
privacy-invasive inferences
despite having access to only a subset of a user's data.
We see the experience significantly impacted
participant willingness to 
share permissions data moving forward. 
Based on the observed privacy violations,
we suggest changes to how smartphone OSes
should gather user consent for data access,
to better inform users about downstream data usage capability.
\end{abstract}

\section{Introduction}
\label{sec:intro}
\input{paper/1-final-introduction}

\section{Background}
\label{sec:background}
\input{paper/2-final-background}

\section{\system System Design}
\label{sec:sysdesign}
\input{paper/3-sysdesign-v2}

\section{Study Methodology}
\label{sec:methodology}
\input{paper/4-methodology-v2}

\section{Results}
\label{sec:results}
\input{paper/5-results}

\section{Recommendations}
\input{paper/6-recommendations-new}

\label{sec:recommendations}

\section{Conclusion}
\label{sec:conclusion}
\input{paper/8-conclusion}

\cleardoublepage
\appendix
\section*{Ethical Considerations}
\label{sec:ethical}
\input{paper/10-ethical-v2}

\section*{Open Science}
\label{sec:osf}
\input{paper/11-osf}

\bibliographystyle{plainurl}
\bibliography{refs}

\appendix
\label{sec:appendix}
\input{paper/9-appendix}

\end{document}

%% file: paper/1-final-introduction.tex
For decades,
online applications have adapted their behavior
by making \textit{inferences} about application users.
By generating hypotheses about a specific user's
preferences and demographic information,
an application can optimize its interactions
with that particular user.
Users now expect personalized experiences,
and companies want to provide those experiences~\cite{mcKinseyPersonalization};
however,
application inferences about a user
may speculate about sensitive information
like a user's health conditions,
political beliefs,
or physical location.
Even users that desire personalized experiences
may nonetheless want to keep aspects of themselves
hidden from online services.
Thus,
understanding the kinds of inferences software can make
is critical for understanding
how software might violate a user's privacy.

For example,
consider online advertising,
a multi-billion dollar industry~\cite{groupm2023adforecast}.
The ecosystem is complex,
but at a high level,
each user click on an ad
will generate revenue for a publisher and an ad exchange,
and will generate site traffic for an advertiser.
So,
publishers, advertisers, and exchanges
are financially incentivized to ``target'' ads,
such that a user is preferentially shown ads
that are likely to be relevant to the user.
Targeting is implemented in practice
by publishers, advertisers, exchanges,
and third-party data brokers
collecting vast amounts of data about
a user's activity
within and across different sites and apps~\cite{englehardt2016online,libert2015exposing},
with the hope that a user's interests
can be inferred via data mining of
the user's information.

Users find this ecosystem problematic
for many reasons.
For example,
users are often surprised and upset
by the scale and granularity
of the data collection~\cite{pew2019privacy,mcdonald2010americans}.
Furthermore,
the massive scale also means that
a single data breach
(or a single unscrupulous data intermediary)
puts sensitive data for millions of people at risk.
These concerns are not hypothetical.
For example,
in 2024,
the FTC sanctioned Gravy Analytics
for collecting and selling user location data
even if users had not provided consent;
the data, associated with roughly 1 billion devices,
included information about user visits
to medical offices, places of worship,
and other sensitive locations~\cite{ftcGravy2024,ftcGravy2025}.
Later, in 2025,
Gravy Analytics was the victim of a massive data breach,
with hackers stealing terabytes of location data
gathered from thousands of popular apps like
Candy Crush, Tinder, Call of Duty, and
several pregnancy tracking apps~\cite{whittaker2025,gravyDataBreach,br2025,cox2025}.
These privacy risks are not unique
to Gravy Analytics---for example, see
Datamaster's unauthorized collection
of the names, physical addresses,
and contact information
for millions of people
afflicted with specific medical conditions~\cite{cppa2026}.

The rapidly improving capabilities
of machine learning models
put user privacy at even greater risk.
Historically,
inference-making about users relied on the analysis
of highly-structured text data
like website profiles,
purchasing histories,
IP addresses,
and cross-site browsing records.
However,
LLMs can now analyze rich media formats
like images, videos, and sound recordings,
generating structured summaries
that can serve as inputs to downstream analyses.
A user's smartphone is a repository
of \textit{user-specific} rich media data
like photos;
a user's smartphone is also a repository
of user-specific, highly-structured text data
like contact lists, calendar events,
and GPS locations.
A smartphone app which receives user permission
to access this repository
can feed the data to an LLM
(located on a server or on the local device)
to extract powerful, privacy-sensitive insights
about a user.
This problem exists
\textit{even if a smartphone app
does not integrate with a legacy back-end system
for data collation}:
a treasure trove of user data has already been collated
by nature of it all existing on the user's device.
As smartphone vendors
begin to restrict how apps can access
unique per-user identifiers
critical to user targeting~\cite{appleAppTrackingTransparency,googleAdId},
mined inferences from data disclosed
\textit{by the users themselves}
become even more important.\footnote{As Meta discussed
in their 2025 Form 10-K submission to the
U.S. Securities and Exchange Commission,
``reduced availability of data signals
used by our ad targeting and measurement tools''
is a significant risk
to Meta's adtech business~\cite{fb2025-10-K}.}
To the best of our knowledge,
there is no prior literature
which describes the ways in which
state-of-the-art LLMs can leverage
user-granted data access permissions
to infer sensitive information about users.

\noindent In this paper,
we provide three contributions:
\begin{itemize}
    \item We provide a new system, \system,
          which implements
          an LLM-based data inference pipeline.
          The \system front-end is a smartphone app
          that requests various permissions from a user
          (\textbf{photo} access, \textbf{calendar/reminders} access, \textbf{location} access,
          and \textbf{contacts} access).
          Leveraging whatever permissions are granted,
          the front-end collects local, user-specific data
          and sends it to an LLM for analysis;
          \system's prompts ask the LLM to
          (1) speculate about sensitive user characteristics,
          and (2) explain \textit{why} those inferences
          were made.
    \item We conduct an IRB-approved user study
          with 465 participants to catalog
          what kinds of inferences \system can make,
          the extent to which those inferences are accurate.
          Our high-level observation is that,
          with access to only a subset of
          a user's smartphone data, \system can often make
          invasive, accurate inferences about
          the personal lives of a user and
          the people in that user's social circle.
    \item We additionally explore the extent to which viewing
          the inferences \system generated about them
          impacts user willingness to share their data
          moving forward, observing statistically significant
          changes in reported sharing decisions.
          We additionally examine the impact of various
          inference features on participant comfort, focusing on 
          the impact of inference sensitivity and accuracy. 
    \item We use the results of the study
          to recommend changes in how smartphone OSes
          gather consent for data access.
          For example, we suggest changes to how
          smartphone OSes gather consent for data access, such that users are explicitly warned
          on an ongoing basis about how
          \textit{overt} data sharing of (say) photos
          can lead to \textit{implicit} sharing of
          personal information inferrable from those photos.
          We also recommend changes to
          the ``acceptable use'' policies
          for LLMs that \system-style systems
          are likely to use.
\end{itemize}
Importantly,
the empirical results from our user study
\textit{understate} the privacy risks
associated with LLM-based inferences.
In our \system prototype,
LLMs run on a handful of GPUs with limited VRAM,
but real-life data brokers will run their analyses
on much larger clusters;
thus,
we expect that real-life privacy risks 
are worse than the ones
that we describe in this paper.

%% file: paper/2-final-background.tex
\subsection{Online Data Sharing}
\label{sec:targetedAds}
A rich online ecosystem exists
to collect, share, and analyze personal user data.
Some participants in the ecosystem are
\textit{primary data sources},
collecting information directly from users
through participant-run websites, mobile applications,
wearable devices, and
similar client-side vantage points.
Other participants act as \textit{intermediaries};
for example,
\textit{data brokers}
aggregate data from
public government documents (e.g., marriage records),
social media sites (e.g., that contain public user profiles),
and other sources,
enriching that data and selling it~\cite{FTC2014databrokers}.
\textit{Buyers}
purchase user data from primary sources or intermediaries,
using that information to target advertisements~\cite{zibler2024facebook, cyphers2020google},
determine eligibility for public benefits~\cite{epicDC},
surreptitiously track
the financial and medical histories
of active-duty military personnel~\cite{sherman2023},
and so on.

Primary data sources,
intermediaries,
and buyers are all motivated
to perform \textit{inferences} on user data.
Inferences enable the surfacing of user traits
that are not directly obvious from a user's data,
but nevertheless may allow a company
to better characterize a user.
For example,
a user who frequently purchases
prenatal vitamins and maternity clothing
might be pregnant
(and thus might be interested in
social media posts about pregnancy);
a user whose IP address and loyalty card records
are linked to a high-income ZIP code
might be a good target for ads about luxury goods.

\subsection{Permissions Systems}
\label{sec:perms-systems}
On both Android phones and iPhones,
apps execute within a sandbox that
tightly restricts how apps can interact
with the rest of the phone~\cite{androidPlatformSecurityModel,applePlatformSecurity}.
If an app desires access to ``sensitive'' data
(e.g., a user's photos or GPS location),
the app must prompt for consent
via an OS-implemented interface;
the interface must be implemented by the OS
because the interface is part of
the phone's trusted computing base.
The OS vendor decides
how the prompting interface works
and which types of phone data
are gatekept via the interface.

\paragraph{Android:}
Android employs a modified
``ask on first use'' (AoFU) policy
for prompting~\cite{androidSecurityPaper2024}.
The first time that an app wants to access
a protected resource,
the app must request user consent.
The user is given three options: (1) disallow the access;
(2) only permit the access once,
meaning that, if the app desires
access in the future,
the app must gather consent again;
(3) grant the access,
such that the app can access the resource
without additional prompting in the future.
The Android-provided language
in a consent screen
only mentions which resources
an app would like to access---the language
does not explain \textit{why} an app
is interested in those resources~\cite{androidPermissions}.
Android does not force an app
to provide such a rationale
within the app experience itself.
Android documentation \textit{recommends} that,
if a user has previously denied
an access request from an app
but the app wants to issue the request again,
the app should first display
an ``educational UI'' which explains why
the app would benefit from data access~\cite{androidPermissions}.
For an app that wants access to
a phone's location, camera, or microphone,
Android also provides an \textit{optional} way
for the app developer to
specify a rationale that will be visible
in certain system-wide settings screens
like Android's Privacy Dashboard~\cite{androidPrivacyDashboard}.

The Google-run app store (called the Play Store)
requires a developer to explain
what data the developer's app collects,
how that data is used,
and  whether the data is sent off-device
(and possibly shared with third-parties)~\cite{playConsoleHelp}.
Google reviews the declaration at app review time,
and displays the declaration
in an app's user-facing listing on the Play Store.
However,
as Google states~\cite{playConsoleHelp},
Google ``cannot make determinations on behalf of the developers of how they handle user data. Only you possess all the information required to complete the Data safety form.''
In other words,
because Google is not privy
to the details of an app's backend infrastructure,
Google relies on developers to
faithfully explain how an app's
backend-mediated data sharing and data analysis works. 

\paragraph{iOS:}
Like Android,
iOS also gatekeeps sensitive data
behind a two-choice AoFS policy
(``deny'' or ``allow''),
with location data having a third choice
of ``allow once''~\cite{appleProtectedResources}.
Unlike Android,
iOS forces an app developer
to provide a rationale string
for each kind of sensitive data access,
with iOS showing the rationale
within iOS-generated consent prompts;
the review process for Apple's first party app store
will reject apps that
do not specify rationale strings~\cite{appleReviewGuidelines}.
Similar to the Play Store,
the Apple store requires app developers
to submit data privacy policies
that describe what data an app collects
and how that data is shared with third parties~\cite{appleReviewGuidelines}.
These policies are checked during app review
and are displayed in each app's listing
in the Apple store~\cite{applePrivacyInfo,appleReviewGuidelines}.
However,
as with the Play Store,
the Apple Store has no way to verify
how an app's purported backend data handling
is actually implemented.

\subsection{Related Work}

Recent work has shown that,
when a complex model interacts with a user,
the model implicitly generates internal hypotheses
about various user characteristics;
such biases affect how the model will interact
with different users who have the same goals~\cite{chen2024designing, viegas2023system}.
In our paper,
we also explicitly expose model hypotheses,
but focus on scenarios in which a model's \textit{specific goal}
is to infer sensitive characteristics about a user.

A substantial literature investigates
whether smartphone permission systems
enable users to understand the implications
of \textit{overt} data sharing with apps~\cite{felt2012android, cao2021large,balash2024would,prange2024not,bohme2011security, lin2014modeling, wijesekera2015android}.
For example,
Lin et al. demonstrated that a small number
of data access policies
were sufficient to capture
how many users would want to share overt data
with various kinds of apps~\cite{lin2014modeling}. 
Other work studied whether users
fully comprehend the OS-managed permissions interface
and know that, e.g., permissions that are granted
can later be revoked~\cite{prange2024not}.
This prior work
once again focused on how users think about overt data sharing.
This work did not explore whether users know that, e.g.,
sharing calendar events with an app
can allow the app to infer a user's health status
or demographic information.

Reitinger et al. used a browser extension to
(1) visualize a user's Google Ad Settings, and then
(2) prompt the user for their discomfort level
with the amount of behavioral tracking that Google Ads
was performing~\cite{reitinger2024does}.
This work focused on exposing coarse-grained user profiling;
in contrast,
our work focuses on the fine-grained
behavioral and demographic inferences that LLMs can produce.
Our user study also analyzes a fuller set of
user attitudes towards data sharing and its privacy implications.

%% file: paper/3-sysdesign-v2.tex
\begin{figure*}[t] 
    \includegraphics[width=0.9\textwidth]{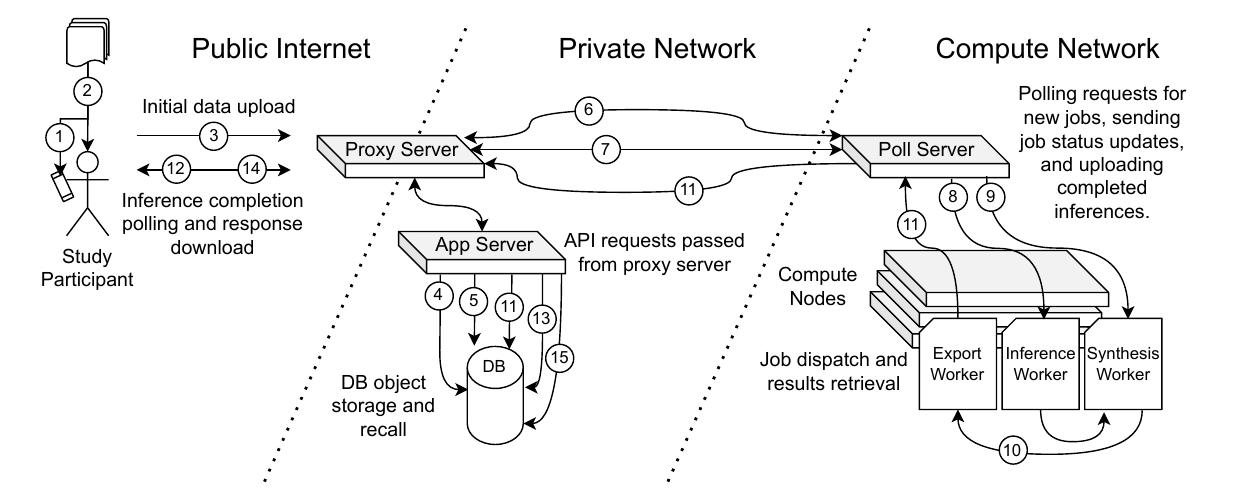}
    \caption{\textbf{\system System Architecture} The user first grants permissions to the client-facing smartphone application \Circled{1}, which then filters and aggregates the user’s data \Circled{2}. Using the public API, the application uploads the data \Circled{3}, which is then stored in the backend database \Circled{4}. The user’s unique session ID (uSID) is marked with the state \texttt{USER\_DATA\_SUBMITTED} \Circled{5}. The polling worker retrieves uSIDs marked \texttt{USER\_DATA\_SUBMITTED} and passes them to the session worker \Circled{6}. The session worker retrieves the corresponding data \Circled{7} and submits it to both the inference worker \Circled{8} and the synthesis worker \Circled{9}, after which the uSID is marked \texttt{JOB\_SUBMITTED}. The resulting inferences are then sent to the export worker \Circled{10}, which marks the uSID with the state \texttt{JOB\_COMPLETE} \Circled{11}. The client-facing smartphone application polls the backend until the uSID is marked \texttt{JOB\_COMPLETE}, at which point the app downloads the inferences, presents them to the user as an in-app survey \Circled{12}, and marks the uSID as \texttt{SURVEY\_STARTED} \Circled{13}. The user completes the survey, the app uploads the survey responses \Circled{14}, and the uSID is finally marked \texttt{SURVEY\_COMPLETE} \Circled{15}.}
    \label{fig:sys_design}
\end{figure*}

In the wild,
the legacy adtech ecosystem already
collects, analyzes, and shares
sensitive user data in troubling ways.
To examine the additional harms
posed by LLM-driven data analysis,
we built \system.
\system collects user data on a smartphone,
preprocesses that data on the client-side,
and then uploads the data to server-side LLMs.
The LLMs try to profile a user,
generating inferences that
we believe are representative of the ones
that real-life online service will want to generate.
However (and importantly),
we designed \system to
(1) center user choice, and
(2) operate in a transparent way.
For example,
users determine what data \system receives,
and users get to see both
the inferences that \system generates
and the reasons why \system's models
claim to have generated those inferences.
\system encrypts a user's data
in transit and at rest.
Furthermore,
once users finish their engagement with \system,
\system deletes a user's raw data
and all its by-products (including
any inferences that \system made
by inspecting user data).
\system therefore allows us
to explore the privacy risks of LLM-mediated data mining
without exploiting users
or putting their data (raw or derived) at risk.

\system consists of three components:
a client-facing smartphone application,
a server-side application and database,
and a server-side compute infrastructure
for generating inferences.
Figure~\ref{fig:sys_design}
provides an overview of the end-to-end architecture.
We provide more detail below.

\subsection{User-Facing App}
\label{sec:harvestApp}
The user-facing app
receives data access permissions from a user,
sends the accessible data to remote servers
for inference mining,
and displays the results to the user.
In this section,
we focus on the implementation of the app;
see Section \ref{subsec:app-user}
for a discussion of the user-facing experience.

The application
requests permission to access
five kinds of smartphone data:
contacts, GPS location, photos, calendar events,
and (on iPhones only) reminders.
If the user grants permission,
the app creates a JSON bundle
which contains the accessible data.
We limit the size of the bundle
for two reasons:
to prevent slow bundle uploads
to the LLM servers,
and to reduce the amount of data
that our modestly-scaled GPU cluster must analyze
(\S\ref{sec:industryRepresentativeness}).
Our app uploads at most
5,000 calendar+reminder events,
5,000 contacts,
the current location (not also historical locations),
and only 10 photos.
The exported JSON bundle is stored in memory
and deleted once the app has completed the bundle upload.

\paragraph{Image filtering:}
\label{image-filtering}
Most user phones will contain more than 10 photos.
Given a limited bundle size,
the \system app employs a variety of heuristics
to select 10 photos that
are likely to contain sensitive information about a user.
For example,
the app preferentially selects photos from
the ``Screenshot'' and ``Favorites'' directories,
as well as from directories whose names
contain keywords like ``private'' or ``secret.''
The app also prefers photos which
embed GPS locations;
the app infers that the most frequently embedded location
is the user's probable ``home,''
and that locations which
are furthest from the home location
represent vacations or other
potentially interesting life events.
Once the final 10 photos are selected,
images categorized as
receipts or identity documents are sharpened
to improve an LLM's ability to read the text.
The photos are then compressed
before being added to the JSON bundle. 

\paragraph{Encryption:}
All network communication between
the \system app and the \system backend
is encrypted via HTTPS.
Inference data fetched from the backend
is stored on the client-side
encrypted via iOS's built-in AES-256 full disk encryption~\cite{apple2026dataprotection},
or a \system-specific SQLCipher\cite{sqlcipher}
encrypted database (also using AES-256) on Android.

\subsection{Data Ingestion and Storage}
The client-side app submits raw user data
and retrieves inference results
via HTTPS-encrypted communication
with the \system proxy server.
The proxy server
is the only part of the \system system
that is exposed to the public internet.
The proxy’s functionality is limited to
request forwarding to
\system's application server;
due to privilege separation~\cite{privSep},
the proxy server cannot run application logic
or access backend databases.
The proxy acts as a protective layer, hiding internal details of the application server from direct exposure to the internet. It can filter malicious traffic, block DDoS attacks, and act as a Web Application Firewall (WAF).
The proxy only exposes the minimal network ports required for HTTPS traffic,
limiting the proxy's threat surface.

The application server
only accepts requests from the proxy.
The application server
is responsible for validating incoming data
and writing it to a backend relational database.
The database server only accepts requests from the application server, is isolated from the public internet, and all participant data is encrypted at rest.

For each batch of submitted JSON data,
the application server
generates a unique session ID (uSID)
which is used to coordinate operations
between the application server, database, and compute environment.
Once a batch is validated and stored,
the application server
marks the corresponding uSID
as ready for processing.

\subsection{Secure Analysis Environment}
\label{sec:secureAnalysis}
\system's inference generation
occurs within a university-managed
secure computation cluster
that was specifically designed to handle sensitive datasets.
\system's job poller process,
running inside the cluster,
periodically queries the application server
(via the proxy server)
to discover uSIDs that
have been flagged as ``ready for processing.''
When such a uSID is identified,
the job poller
dispatches a work item to a queue,
causing several worker processes to spawn:
a session worker fetches the session's batch data,
an inference worker
mines the batch data for interesting observations,
and a synthesis worker
examines all of the generated inferences
and identifies a subset that
maximize a scoring function.
We describe the inference worker
and the synthesis worker in more detail below.
For now,
we observe that
all of the data processing occurs
within the secure compute environment,
and is done in accordance with
institutional policies for handling sensitive data.

\paragraph{Inference worker:}
Upon receiving batch data,
an inference worker first reformats the data
to improve the likelihood
of successful downstream inferences
and/or reduce the amount of data
which must be analyzed at inference time.
For example,
the worker down-samples the contact list metadata,
keeping only the area code, name, and
organization name for each contact.
The reason for keeping organization names is that
we empirically observed that
an organization name
is often used to denote
the context of a relationship
(e.g., John Smith, Tinder).
The worker also down-samples
dates in calendar events,
keeping the year, month, day, hour, and minute,
but removing the seconds.
The inference worker uses
the Pelias geocoder~\cite{pelias} to
convert raw GPS coordinates into human-readable addresses.

Once the data cleaning is finished,
the inference worker uses
llama-3.3-70b to analyze
calendar events, calendar reminders,
contact lists,
and location data.
The inference worker uses
llama-3.2-vision-90b to inspect photos.
\system directs both models
via \system-specific prompts which
ask the models to make inferences
about sensitive topics like a user's
politics, social relationships,
medical diagnoses, and behavioral tendencies.
These initial inferences are generated in parallel,
split by permission type. 

\paragraph{Synthesis worker:}
When all inference workers are complete, 
draft inferences are handed off
to a synthesis worker.
\system prompts the synthesis worker
to identify the subset of prompts
that are most likely to exhibit
sensitivity, specificity, and accuracy.
We spent significant time
tuning the inference and synthesis worker prompts; however,
for ethical reasons,
we do not include the raw prompts
in this paper. 
For additional discussion of this topic,
see Section~\ref{sec:ethical}.

Once the final inferences have been chosen,
the synthesis worker forwards them to the proxy,
who then asks the app server
to store the inferences in the
encrypted backend database.
The app server
updates the batch flag for the uSID
to indicate that inference data is now available;
at this point,
the app server deletes the raw user data
stored in the batch.
The client-side \system app
detects the updated batch flag
and retrieves the encrypted inferences
via HTTPS.
Once the client has downloaded the inferences,
the app server deletes them
from the backend database.
Meanwhile,
on the client,
the user completes the survey workflow
described in Section~\ref{sec:methodology}.

\subsection{Industry Representativeness}
\label{sec:industryRepresentativeness}
We built our \system prototype
to explore the privacy harms
enabled by LLM-generated inferences.
As we discuss in Section~\ref{sec:results},
our prototype generated inferences that
were more accurate than those made possible
via traditional adtech tracking mechanisms.
However,
we expect that real-life online services
will be able to generate even more powerful inferences.
For example,
\system uses off-the-shelf, open-source models
which were not fine-tuned
for targeted advertising use cases; 
commercial entities are capable of
training much more sophisticated models.
Also,
relative to a company like Google or Meta,
our compute resources and data resources were extremely limited.
Our cluster
had 8 H200 GPUs,
and we needed to restrict our per-batch compute time
to three minutes to ensure that
participants in the user study (\S\ref{sec:methodology})
would not abandon the study.
Our lack of compute power and compute time
forced us to limit the size of each batch,
e.g., by restricting the number of uploaded photos
to 10 (\S\ref{sec:harvestApp}).
Technology companies like Meta
have tens of thousands of GPUs,
as well as the luxury of being able to
asynchronously stream new user data
to those GPUs
without concern for user drop-out.
Thus,
the capabilities of our \system prototype
represent a lower bound on
industrial-strength inference capabilities.

%% file: paper/4-methodology-v2.tex
To evaluate the real-world privacy risks
of LLM-generated inferences,
we conducted a large-scale, IRB-approved user study,
deploying the \system client-side app
on real user phones.
The high-level goals of the study were to
(1) empirically catalog the types of inferences
that LLMs can generate,
and (2) ask users to respond to those inferences
e.g., with respect to whether the inferences were accurate,
and whether users,
after being shown what LLMs can infer,
would reconsider granting data access to smartphone apps.

\subsection{Recruitment Procedure}
\label{subsec:recruitment}
We recruited participants
on Prolific, a well-known recruitment platform
for user study research~\cite{prolific}. 
We required that participants
have a smartphone,
be located in the U.S., and
be at least 18 years old.
We recruited Android and iOS smartphone owners equally. 
As part of the advertisement and onboarding process,
we provided users with detailed information
about \system's responsible use of sensitive data;
for example,
we provided links to readable privacy policies,
and to a detailed overview of the study.
We provide more context
on the ethics of our recruitment process
in Section~\ref{sec:ethical}.
Participants were paid \$15 for their participation,
and took on average 33 minutes to complete the study.

\subsection{User-facing Workflow}
\label{subsec:app-user}

\begin{figure}
    \centering
    \includegraphics[width=.85\linewidth]{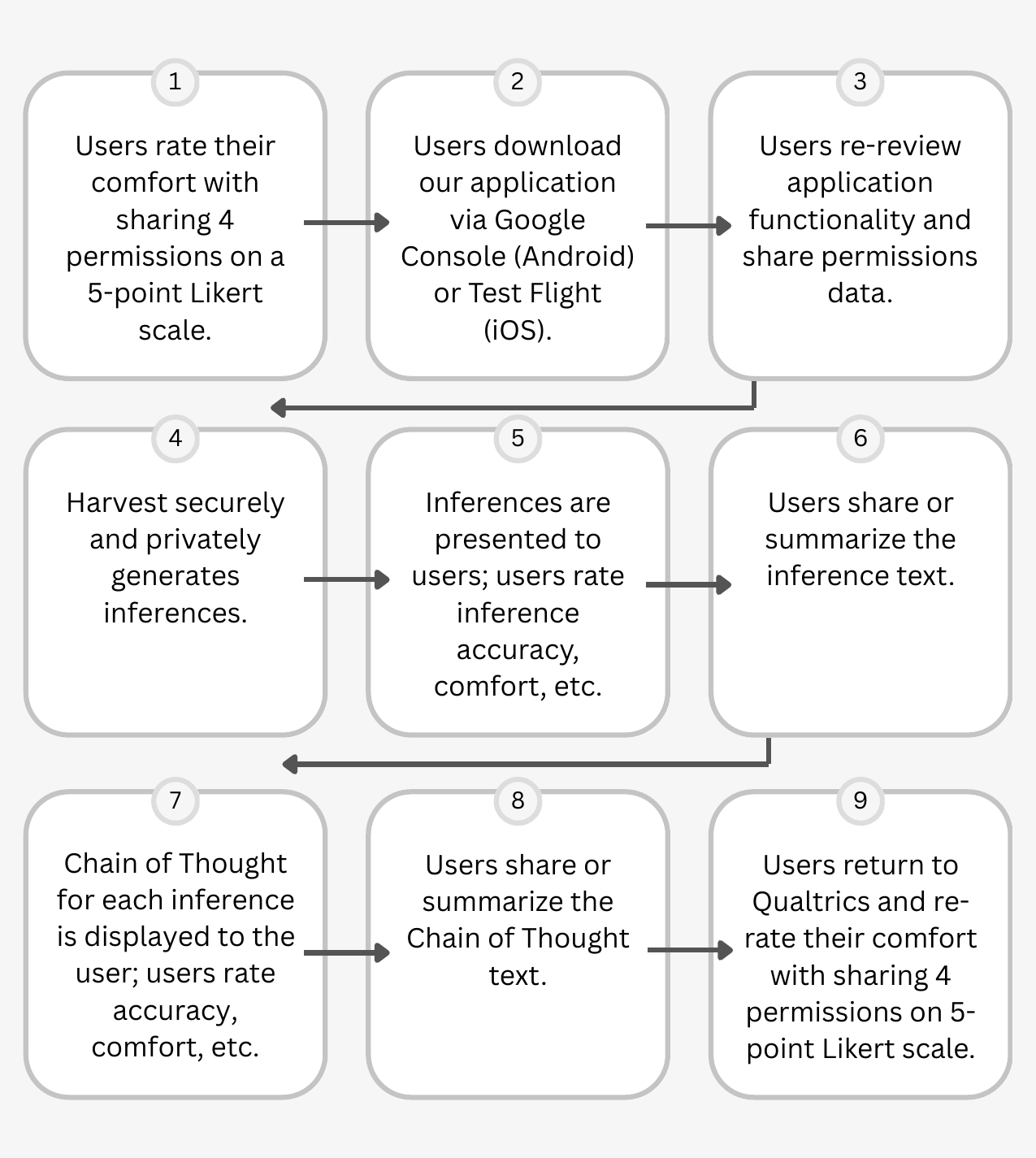}
    \vspace{-.5cm}
    \caption{A high level overview of our study flow.}
    \label{fig:study-flow}
\end{figure}

An overview of our study is shown in Figure~\ref{fig:study-flow}.
We describe each step in detail below. 

\noindent
\textbf{(1) Pre-Intervention beliefs:}
After users signed up on the Prolific website,
the site directed them
to a Qualtrics survey form,
where users re-confirmed that
they met the participant criteria and
wanted to participate in the study.
We then asked users
how comfortable they would be
to share four different data types
(calendar, contacts, photos, and location)
with a social media company;
to provide users with real-life context for the question,
we showed them
an image of a typical Android permissions pop-up.
Participants rated their comfort
on a 5-point Likert scale. 

\noindent
\textbf{(2) Application download:}
Next,
the Qualtrics page
showed a video showing how users
could install the iOS or Android version
of the \system app
via TestFlight on iOS or Google Console on Android.
The video also
described the study's goals and privacy policies again,
and emphasized that \system-generated inferences
might involve sensitive information
and might be inaccurate.
Appendix~\ref{app-screenshots} provides
UI screenshots for this part of the workflow.

\noindent
\textbf{(3) Access requests:}
After a user has installed and launched
the \system app,
the app again displays
an overview of the study's goals and privacy policies
(see Appendix~\ref{app-screenshots}).
The app then requested permission
to access a user's photos, location,
contact list, calendar events,
and (on iOS) reminders.
These permissions are among the most common requested by popular social media applications (see Appendix~\ref{companies-collect}).
The app required
at least partial granting of each permission type
(e.g., the app required access to
at least 10 photos and 30 contacts).

\noindent
\textbf{(4) Inference generation:}
After the participant
granted the necessary permissions,
the app displayed
a loading screen which indicated that
\system was generating inferences.
The \system backend generates five inferences to show to the user.
On average,
\system required 2.73 minutes to complete this step.

\noindent
\textbf{(5 \& 6) Inference presentation and user responses:}
When the inferences
were received by the participant's device,
the \system app
displayed each inference one at a time.
The app asked
questions
that measured the participant's comfort and surprise
with each inference.
Participants were also asked to evaluate
the accuracy of each inference.
All of these questions received responses
via a 5 point Likert scale.
The apps also asked users
to provide open-ended responses
describing what made inferences (in)accurate,
and why users had selected their chosen comfort rating.
The full set of study questions
is enumerated in Appendix~\ref{survey-questions}.

For each inference,
users could then choose to share (1) the raw LLM-generated inference text or (2) a user-generated summary of the LLM output (providing detail to the degree they felt comfortable)
The app explicitly requested participants
not use the first option
if inference text contained
personally identifiable information
like names or street addresses. 
\\
\textbf{(7-8) Reasoning presentation and user responses:
}
After answering questions about all five inferences,
users were presented with
the chain of thought used by the model
to produce each inference. 
We provide sample chains of thought
(and a fuller analysis of their content)
in Section~\ref{subsubsec:RQ1}. 
For each chain of thought,
the app requested the user's comfort and surprise
with the model-provided reasoning,
and asked the user to evaluate the accuracy of the chain of thought.
The app additionally requested open-ended explanations
for both the user's comfort and the chain of thought's accuracy.

\noindent
\textbf{(9) Post-intervention beliefs:}
The app then directed users
back to the Qualtrics site.
The site prompted users to enter a completion code
(needed for a participant to receive full payment).
The site then provided
instructions for deleting the \system app.
Once a user deleted the app,
the survey repeated the questions from Step 1,
asking a user to (re-)evaluate their willingness
to grant various permissions with a smartphone app.
The survey prompted the user for
an open-ended response about how the study
changed their attitudes about data sharing.
Next,
the survey asked users to report
demographic information like
age and highest level of education.
Finally,
the user complete an IUIPC survey~\cite{gross2021validity}, 
a standard framework for assessing participant privacy concern. 

\subsection{Analysis Methodology}
To evaluate \system,
we used both inferential and descriptive statistical methods
to evaluate close-ended participant response data.
We provide low-level details
of our statistical modeling
in Appendix~\ref{subsec:app-stats-modeling}.

\subsubsection{Statistical Analysis}
\label{subsubsec:stats}

\textbf{Intervention-Level Ordinal Mixed-Effects Model:}
We evaluate if use of the \system application influences user decision-making for each permission type.
Participants report their comfort with sharing 4 different permissions, once before app use, and again after app use. 
We modeled the change in self-reported comfort as an ordered outcome. 
We incorporated intervention-specific characteristics derived from participants’ interactions with \system: specifically self-reported accuracy and comfort with system-generated inferences. 
Inference-level responses were aggregated at the participant level by summing accuracy and comfort ratings across all inferences, capturing participants’ cumulative exposure to and perceived quality of \system's outputs.
To account for repeated measurements, data was reshaped into long format with one observation per participant per time point. 
Comfort with sharing permissions data was treated as an ordered categorical outcome. 
Aggregated accuracy and comfort predictors were standardized (z-scored) to improve model stability. 
Because the outcome reflects ordered categories, we used an ordinal regression framework rather than linear models.

We fit a cumulative link mixed-effects model in R using the \texttt{clmm} function from the \texttt{ordinal} package with a logit link. 
We fit separate CLMMs for each permission type.
Fixed effects included time (pre/post), standardized summed inference accuracy, standardized summed inference comfort, and their interactions. 
A random intercept for participant accounted for individual differences in baseline comfort. 
Results are reported as log-odds of higher comfort categories, enabling estimation of predicted probabilities for each level of permission sharing. Results are discussed in Section~\ref{sec:results}, with greater implementation details in Appendix~\ref{subsec:app-stats-modeling}.\\
$ $\\
\noindent
\textbf{Correlation of Ordinal Variables:}
Our analysis focused on the accuracy and user comfort values associated with individual inferences.
To analyze the relationship between accuracy and comfort, we used Spearman's rank correlation, which is a non-parametric measure for the monotonic relationship between two variables. 
We additionally stratified the data across inference data types~\cite{gross2021validity}. 
We reported coefficients and significance values, as well as the direction and strength of the relationship (See Appendix~\ref{subsec:corr-analysis}).
We used a descriptive correlation analysis to prevent mathematical coupling between inference-level accuracy and comfort values in our mixed-effects models~\cite{squara2008mathematic}. 

\subsubsection{Qualitative Analysis}
\label{subsubsec:qual}
We implemented iterative open coding to analyze participant responses for open-ended questions surrounding comfort, accuracy, and intervention impact~\cite{strauss1990basics}.
To begin, two members of the research team developed an initial codebook, evaluating the responses of 30 participants (150 inferences, 150 reasonings).
The researchers then independently coded the remaining responses in rounds of 30 participants/150 inferences/150 reasonings, evaluating Krippendorff's $\alpha$ per section at each round, addressing disagreements, and making adjustments to the codebook when necessary. 
After three rounds, the researchers reached a Krippendorff's $\alpha$ of 0.8 across the codebook~\cite{krippendorff2011computing}. 
The two researchers evenly divided and coded the remaining participant responses.

\subsection{Study Limitations}
\label{subsec:limitations}

We speculated that,
relative to the general population,
our recruited participants
might be less concerned with privacy
due to their willingness to
share personal data with us. 
To address this concern,
we repeatedly emphasized to participants
that the \system system
was designed to safeguard user privacy
at every step;
via this emphasis,
we hoped to encourage
more privacy-sensitive users to participate.
We believe that our approach was effective,
at least as indicated by the many participants
who explicitly stated that
they felt comfortable sharing private data with us.

While our study experienced initial participant attrition (approximately half of eligible individuals recruited via Prolific did not ultimately accept or consent to participate), the average privacy behavior score of our final participant pool, measured using the IUIPC framework, is comparable to those reported in prior work~\cite{balash2021security, colnago2022concern}. 
This suggests that the privacy behavior of our population was not meaningfully impacted by this attrition.

Additionally,
our study structure might have introduced
social desirability bias:
participants might have exaggerated
self-professed privacy concerns
if participants thought that such concerns
were desired by the study organizers.

Our study exhibits sample bias
stemming from our participant population.
Notably, as discussed in Section~\ref{demographics},
our participants were lower income than the national distribution, 
which may impact inferences
about socioeconomic status.

%% file: paper/5-results.tex
\subsection{Participant Demographics}
\label{demographics}

After removing incomplete survey submissions,
our study ended with 465 participants. 
Compared with the US Census Bureau ACS 2019--2023 data,
our population was generally representative.
Our population's race demographics
slightly overrepresented white individuals (68.0\% vs 59.8\%) and
individuals of two or more races (8.6\% vs 4.6\%),
and slightly underrepresented Hispanic or Latino participants (12.3\% vs 20.0\%)~\cite{Census2024ACSDP1Y2024.DP05}.
With respect to age,
our study participants underrepresented individuals
65 and older (8.6\% vs 22.9\%) and therefore
modestly overrepresented all other age groups
above 18~\cite{Census2024ACSST1Y2024.S0101}.
For education,
our study population overrepresented individuals
with a bachelor’s degree or higher (49.7\% vs 34.1\%),
and overrepresented individuals that possess
a high school diploma or higher (99.1\% vs 89.7\%)~\cite{Census2024ACSST1Y2024.S1501}.
Appendix~\ref{appendix-demographics} provides
a full overview of
the self-reported demographics of our study population.

Participants in our study had IUIPC privacy belief scores\footnote{
See the full question set in Appendix \ref{pre-post-qs}.}
that were comparable with those seen in previous work:
our study participants had an average score of 5.78,
with previous work being situated in a similar range~\cite{balash2021security, colnago2022concern}.
This signifies comparable privacy concern to a general population. 
Our participant population
did over-represent low income individuals,
with approximately 28.7\% of our participants
making under \$25k annually
as compared to 8.2\% of the general US population.
This type of skew is expected
when drawing users via online platforms like Prolific;
as we mentioned in Section~\ref{subsec:limitations}, 
the skew might impact
the accuracy of \system's inferences
pertaining to socioeconomic status. 

\subsection{Inference Content and Quality}
In this section,
we use a mixed-methods approach
to explore the inferences
that \system generated about study participants.
We examine common inference subjects,
the chains of thought provided by the LLM,
and the user-perceived accuracy of
both the inferences and the chains of thought. 
We also demonstrate that,
even though \system-generated inferences
are sometimes incorrect,
they are often more correct and more specific than
inferences output by
the traditional adtech ecosystem.

\subsubsection{\textbf{(RQ1): What types of inferences were made by \system?}}
\label{subsubsec:RQ1}

\begin{figure}[t!]
    \centering
    \includegraphics[width=\linewidth]{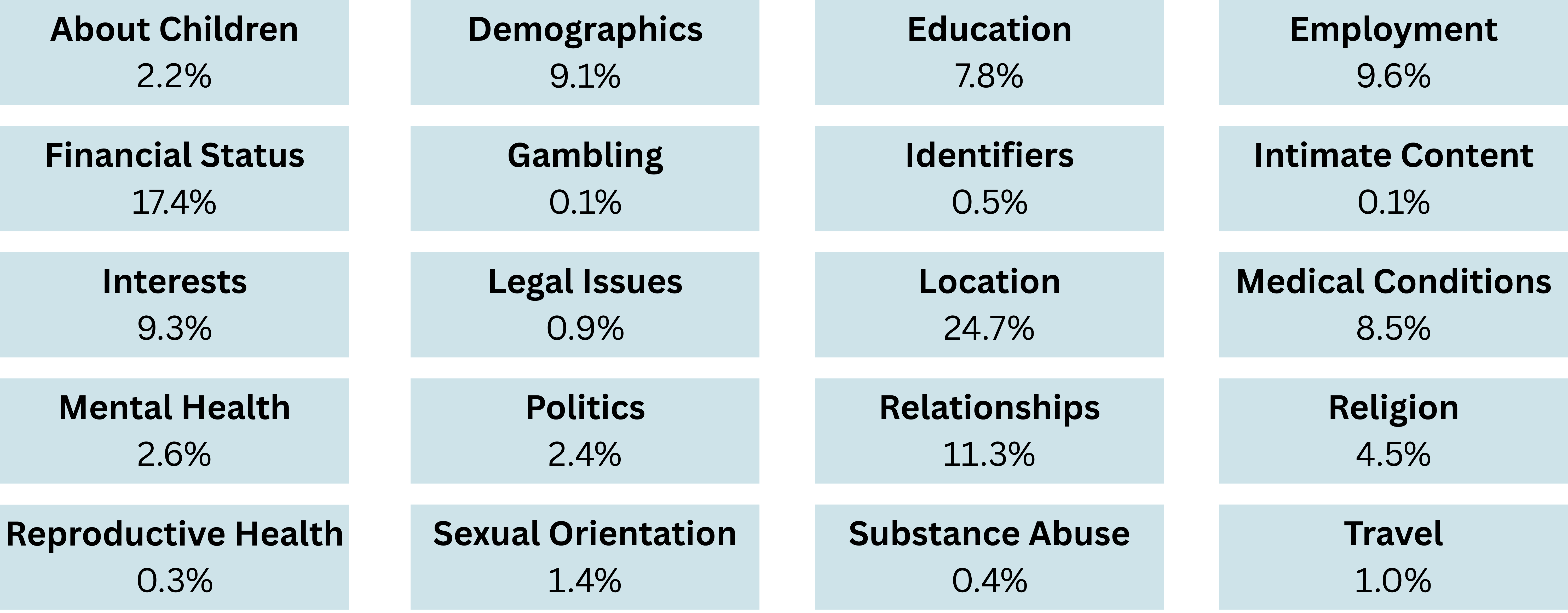}
    \vspace{-.5cm}
    \caption{Categories of inferences produced using \system across a sample of 465 users.}
    \label{fig:inf-types}
\end{figure}

We performed a qualitative analysis
of the generated inference text,
using the method outlined in Section~\ref{subsubsec:qual}
to identify themes.
The range of observed inference types
is shown in Figure~\ref{fig:inf-types}.
We observe that many of these inferences
involve sensitive topics.
For example,
many inferences focused on medical conditions (8.5\%)
and mental health conditions (2.6\%).
The inferences speculated about diagnoses,
doctor names and office names,
and medications taken. 
Below,
we provide examples of health-related inferences that were rated as accurate by the participants: 

\begin{tcolorbox}[
  colback=gray!10,
  colframe=gray!40,
  boxrule=0.2pt,
  left=1pt,
  right=1pt,
  top=1pt,
  bottom=1pt,
  boxsep=0pt
]
\itshape
``You have a history of struggling with attention deficit hyperactivity disorder (ADHD) and are currently taking Concerta to manage your symptoms.''
\end{tcolorbox}

\begin{tcolorbox}[
  colback=gray!10,
  colframe=gray!40,
  boxrule=0.2pt,
  left=1pt,
  right=1pt,
  top=1pt,
  bottom=1pt,
  boxsep=0pt
]
\itshape
``You or someone close to you has a child with a genetic condition and has sought medical attention for this condition.''
\end{tcolorbox}

\begin{tcolorbox}[
  colback=gray!10,
  colframe=gray!40,
  boxrule=0.2pt,
  left=1pt,
  right=1pt,
  top=1pt,
  bottom=1pt,
  boxsep=0pt
]
\itshape
``You or someone close to you has experienced trauma or stress related to memory loss and is seeking information on C-PTSD.''
\end{tcolorbox}

We additionally observe inferences
surrounding reproductive health (0.26\%);
these inferences typically identify plans
to become (or not become) pregnant.

\begin{tcolorbox}[
  colback=gray!10,
  colframe=gray!40,
  boxrule=0.2pt,
  left=1pt,
  right=1pt,
  top=1pt,
  bottom=1pt,
  boxsep=0pt
]
\itshape
``You are a woman who has experienced issues with birth control and is seeking alternative methods.''
\end{tcolorbox}

\begin{tcolorbox}[
  colback=gray!10,
  colframe=gray!40,
  boxrule=0.2pt,
  left=1pt,
  right=1pt,
  top=1pt,
  bottom=1pt,
  boxsep=0pt
]
\itshape
``You are interested in prenatal care or parenting resources, which may indicate that you are preparing for a new addition to your family.''
\end{tcolorbox}

Across categories,
we observe that inferences
do not solely target a phone's user,
but may also reference individuals
who are close to the user. 
Table~\ref{tab:about-who} displays
a taxonomy of target subjects.
The diversity of target subjects is important
because it shows that
when a user decides to share data with an app,
the privacy of the user
\textit{and the user's social circle}
is impacted.

\begin{table}[t]
\centering
\begin{tabular}{lcc}
\toprule
Subject of Inference & \% of Inferences \\
\midrule
    User & 70.3\% \\
    User or Someone Close To Them & 23.9\% \\
    Family or Relationship & 9.6\% \\
    Child & 2.2\% \\
    Someone Else & 0.8\% \\
\bottomrule
\end{tabular}
\caption{Target subject of inferences produced by \system. Note that percentages do not sum to 100\%, as a single inference may reference multiple subjects (e.g., ``You are pregnant, and have a child in elementary school'')}
\label{tab:about-who}
\end{table}

We also observe that inferences vary in \textbf{specificity}. 
We see that some inferences are relatively general,
i.e., they apply to a large anonymity set.
For example:
\begin{tcolorbox}[
  colback=gray!10,
  colframe=gray!40,
  boxrule=0.2pt,
  left=1pt,
  right=1pt,
  top=1pt,
  bottom=1pt,
  boxsep=0pt
]
\itshape
``You are a homeowner with a moderate to high income level and reside in a suburban area, likely with dependents.''
\end{tcolorbox}
Other inferences are more specific,
but still not identifying:
\begin{tcolorbox}[
  colback=gray!10,
  colframe=gray!40,
  boxrule=0.2pt,
  left=1pt,
  right=1pt,
  top=1pt,
  bottom=1pt,
  boxsep=0pt
]
\itshape
``You have placed a bet on a sports game, specifically the [\textit{Redacted: Basketball Team}], indicating an interest in sports betting.''
\end{tcolorbox}
However,
many inferences had small anonymity sets,
containing identifiers, 
such as email or job ID numbers (0.2\%),
exact addresses (1.9\%), or
the specific names of
users, friends, or family members (2.3\%).
For example,
in the context of a user's financial life,
\system inferred an individual's insurance policy number:
\begin{tcolorbox}[
  colback=gray!10,
  colframe=gray!40,
  boxrule=0.2pt,
  left=1pt,
  right=1pt,
  top=1pt,
  bottom=1pt,
  boxsep=0pt
]
\itshape
``You have a car insurance policy with State Farm that is effective from [Redacted: DDMMYYY], [Redacted: DDMMYYYY], and have insured a Chrysler Fiat 500 with a policy number of [Redacted: Policy Number].''
\end{tcolorbox}

We provide a more comprehensive,
de-identified sample of accurate inferences
in Appendix~\ref{example-infs}.

\subsubsection{\textbf{(RQ2): What chain of thought did the model provide?}}
\label{RQ2}
A model's chain of thought
represents a model's self-reported reasoning process
for a particular model output.
We observed that all four data types
(calendar events, photos, location, and contacts)
regularly appeared in chains of thought,
with location being the most popular (36.3\%)
and photos being the least popular (15.1\%). 
As shown by Figure~\ref{fig:appendix-inf-accuracy-by-reasoning-type},
\system produced the most accurate inferences
when chains of thought mentioned
area codes (62.3\% accuracy),
calendar events/reminders (59.3\% accuracy),
contacts (53.9\% accuracy), and
screenshots (52.9\% accuracy).

Sample chain of thought statements demonstrate the wide range of model reasoning capability:

\begin{tcolorbox}[
  colback=gray!10,
  colframe=gray!40,
  boxrule=0.15pt,
  left=1pt,
  right=1pt,
  top=1pt,
  bottom=1pt,
  boxsep=0pt
]
\itshape
``The calendar data lists an event titled ’[Redacted: Name]
Ultrasound’ on [Redacted:DDMMYYYY], which suggests a medical
appointment, possibly related to pregnancy.''
\end{tcolorbox}

\begin{tcolorbox}[
  colback=gray!10,
  colframe=gray!40,
  boxrule=0.15pt,
  left=1pt,
  right=1pt,
  top=1pt,
  bottom=1pt,
  boxsep=0pt
]
\itshape
``The presence of '[Redacted: Name]' in the organization names, which is a term commonly used by the LDS Church, suggests that the user is likely a member of the LDS Church. (contacts source)''
\end{tcolorbox}

\begin{tcolorbox}[
  colback=gray!10,
  colframe=gray!40,
  boxrule=0.1pt,
  left=1pt,
  right=1pt,
  top=1pt,
  bottom=1pt,
  boxsep=0pt
]
\itshape
``The neighborhood's voting history and demographics, as well as the user's location data, suggest that the user is likely to be a Democrat or have liberal-leaning views (location source). Additionally, the neighborhood's reputation for diversity and inclusivity supports the hypothesis about the user's potential involvement in the LGBTQ+ community (location source).''
\end{tcolorbox}

\system's ability to glean information from photos
was particularly interesting.
Using client-side photo selection (\S\ref{image-filtering}), the model found images
containing ID cards (0.2\%),
credit/debit cards (0.2\%),
social media screenshots (1.2\%),
and  text messaging screenshots (1.3\%).
For example:
\begin{tcolorbox}[
  colback=gray!10,
  colframe=gray!40,
  boxrule=0.2pt,
  left=1pt,
  right=1pt,
  top=1pt,
  bottom=1pt,
  boxsep=0pt
]
\itshape
``The image shows a tattoo on your left shoulder, which is visible in the top right corner of the image (image source).''
\end{tcolorbox}
\begin{tcolorbox}[
  colback=gray!10,
  colframe=gray!40,
  boxrule=0.2pt,
  left=1pt,
  right=1pt,
  top=1pt,
  bottom=1pt,
  boxsep=0pt
]
\itshape
``The image shows a screenshot of a sports betting
app, with a list of recent bets and their outcomes, which
suggests you have placed a bet on a recent NFL game''
\end{tcolorbox}
\begin{tcolorbox}[
  colback=gray!10,
  colframe=gray!40,
  boxrule=0.2pt,
  left=1pt,
  right=1pt,
  top=1pt,
  bottom=1pt,
  boxsep=0pt
]
\itshape
``The product's name and description, 'Shoe String King Square Afro Pick with Black Fist', imply a connection to African American culture.''
\end{tcolorbox}

\subsubsection{\textbf{(RQ3): How accurate are inferences?}}
\label{subsubsec:RQ3}
\system was able to generate inferences
which speculated about particular user characteristics---but
how accurate were those speculations?
Table~\ref{tab:inf-acc} shows that
48.8\% of the inferences were rated by users
as ``completely accurate,''
and 16.7\% were rated as ``somewhat accurate.''
To place these results in context,
we note that
the traditional data sharing ecosystem
is already comfortable with imperfect information.
For example,
Neumann et al. observed that, on average,
major data brokers achieved 24.4\% accuracy
on inferences involving gender and age buckets~\cite{neumann2019frontiers}.
We also reiterate
our observation from Section~\ref{sec:industryRepresentativeness}
that our \system prototype
uses orders of magnitude less
data, GPUs, and compute time
than what a real-life, popular online service
would have access to.
Thus,
the accuracies in Table~\ref{tab:inf-acc}
are lower bounds on the accuracies
that we would expect to see in practice.

\begin{table}[t]
\centering
\begin{tabular}{lcc}
\toprule
Accuracy Rating & \% of Inferences \\
\midrule
    Accurate & 48.8\% \\
    Somewhat Accurate & 16.7\% \\
    Neither Accurate Nor Inaccurate & 3.6\% \\
    Somewhat Inaccurate & 13.1\% \\
    Extremely Inaccurate & 17.8\% \\
\bottomrule
\end{tabular}
\caption{Distribution of user-reported inference accuracies}
\label{tab:inf-acc}
\end{table}

\subsubsection{\textbf{(RQ4): What are the causes for inaccuracy?}}
\label{RQ4}

To better understand
why \system sometimes generated inaccurate inferences,
we qualitatively analyze user-reported comments about incorrect inferences.
Table~\ref{tab:inaccuracies} provides a taxonomy
of the user explanations.
We use the term \textit{accurate-adjacent} inferences
to describe those that are relevant to a user,
and valuable to advertisers and data brokers,
but nonetheless contain qualitative errors.
For example,
an inference might not apply to the user,
but to a person who is close to them:
\begin{tcolorbox}[
  colback=gray!10,
  colframe=gray!40,
  boxrule=0.2pt,
  left=1pt,
  right=1pt,
  top=1pt,
  bottom=1pt,
  boxsep=0pt
]
\itshape
``You have a personal connection to a horse, possibly as an owner or frequent rider.''
\end{tcolorbox}
\noindent
\system produced the inference above
after examining photos that depicted the user
standing next to a horse. 
The user explained that this image
was not of them, but their daughter.
We note that these types of ``social proximity'' errors
are present in
the traditional adtech/data mining ecosystem too.
For example, a child might search for toys
on their parents' computer;
the parent did not issue the search,
but the parent may nonethess purchase
the searched-for toy online
in response to a real-life request from the child.

As shown in Table~\ref{tab:inaccuracies},
we refer to the second category of flawed inferences
as \textit{incorrect assumptions}. 
These inferences often arise
due to incorrect generalizations
of a true fact.
For example,
\system might infer that,
because a user lives in a location
that has a high average income,
the user themselves has a high income. 
These kinds of incorrect inferences
also exist in the traditional adtech ecosystem,
e.g., advertisers target users
based on a zipcode inferred from IP geolocation.

The final category of inaccuracy
arises from what we call \textit{incorrect inferences}.
These inferences are fundamentally incorrect,
and often arise from basic flaws in the model's reasoning.
For example,
the model might incorrectly interpret an acronym
that the model extracted from text in photo.
Errors might also be induced
by problems in data preprocessing,
such as incorrect mapping of GPS locations
to human-readable street addresses.

\begin{figure*}[t]
    \centering
    \caption{Reported inference accuracy stratified by reported user comfort across all generated inferences.}
    \includegraphics[width=.75\textwidth]{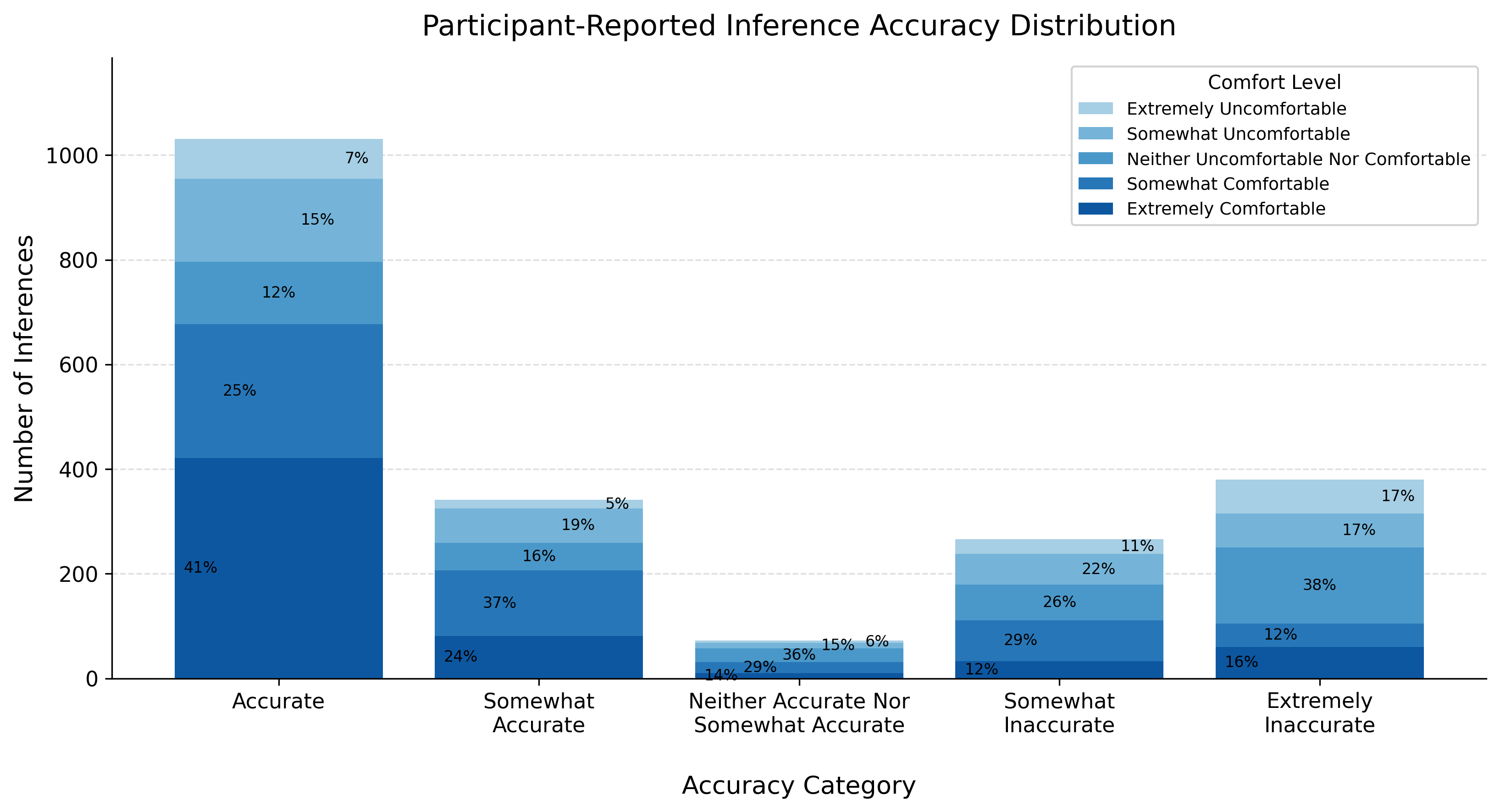}
    \label{fig:acc-dist-table}
\end{figure*}

\begin{table*}[t]
\caption{Popular inaccuracy subtypes for each identified inaccuracy group, summed across inferences and chains of thought. \\ Full list in Appendix Table~\ref{tab:inaccuracies-appendix}}
\label{tab:inaccuracies}
\begin{tabular}{|c|c|c|c|c|}
\hline
\makecell{Inacc. \\ Group} & \makecell{Inacc. \\ Type} & \makecell{\% of\\ Inacc} & Definition & Participant Example \\
\hline
                         \multirow{2}{*}{\makecell{Accurate-\\Adjacent}}  & \makecell{Wrong Date/Time} & 18.7\% & Misinterprets timing & \makecell{``[I] was involved but no longer am.''} \\
                        \cline{2-5}
                         & \makecell{Wrong Person} & 11.9\% & \makecell{Correct but applied to \\ the wrong person.} & \makecell{``My husband has a moderately\\high income. I do not.''} \\
                        \hline
                         \multirow{2}{*}{\makecell{Incorrect\\Assumption}}  & \makecell{Wrong Income\\Assumption} & 33.3\% & \makecell{Assumes incorrect\\financial circumstances} & \makecell{``I'd say my income is\\more middle class.''} \\
                        \cline{2-5}
                         & \makecell{Wrong Demographic\\Assumption} & 8.6\% & \makecell{Assumes incorrect demographic\\(age, race, gender, etc)} & \makecell{``I am 22, so I am slightly\\below the described age bracket.''} \\
                        \hline
                         \multirow{2}{*}{\makecell{Incorrect}}  & \makecell{Incorrect Info Used} & 10.3\% & \makecell{Assumption based on \\ an incorrect fact} & \makecell{``The address is in the city, \\ not suburbs.''} \\
                        \cline{2-5}
                         & \makecell{Incorrect\\ Interpretation} & 4.5\% & \makecell{Incorrect interpretation of \\ acronyms, titles, etc} & \makecell{``It's mixing up [the name of]\\ a test for a relationship.''} \\
\hline
\end{tabular}
\end{table*}

\subsection{Impact of Inferences on Participant Comfort and Decision Making}

In this section, we examine participants' qualitative experiences with \system.
For individual inferences, we examine how accuracy impacted participant comfort.
We then observe the \system experience's impact on users' reported future permissions decisions.

\subsubsection{\textbf{(RQ5): What is the relationship between inference accuracy and participant comfort?}}
\label{sec:rq5}

As discussed in Section~\ref{subsubsec:RQ3}, data brokers often \textit{unintentionally} generate inaccurate inferences, e.g., due to incorrect underlying data or ineffective processing mechanisms.
Additionally, companies may \textit{intentionally} deploy privacy mechanisms which introduce inaccuracies into user data~\cite{google2024sharing, voorhees2025apple, edmund2020how}.

One might expect, as observed by Reitinger et al.~\cite{reitinger2024does}, that participants would find accurate inferences ``creepy,'' and be relieved by incorrect predictions.
However, this was not the case for our participants; with a Spearman correlation coefficient value of 0.263, there existed a weak (but present) monotonic relationship between increased inference accuracy and increased user comfort with those inferences.
Figure \ref{fig:acc-dist-table}
shows the distribution of reported inference accuracies, stratified by the participant's reported comfort for the inference. Participants who saw accurate inferences reported being somewhat or very comfortable 66\% of the time; in contrast, when participants were shown extremely inaccurate inferences, participants were somewhat or very comfortable only 27\% of the time.
Below, we explore the features that impacted user comfort, focusing primarily on the accuracy of the inference. 

\begin{figure*}[t]
    \centering
    \caption{Reported willingness to grant various permissions before and after use of the \system application.}
    \includegraphics[width=\linewidth]{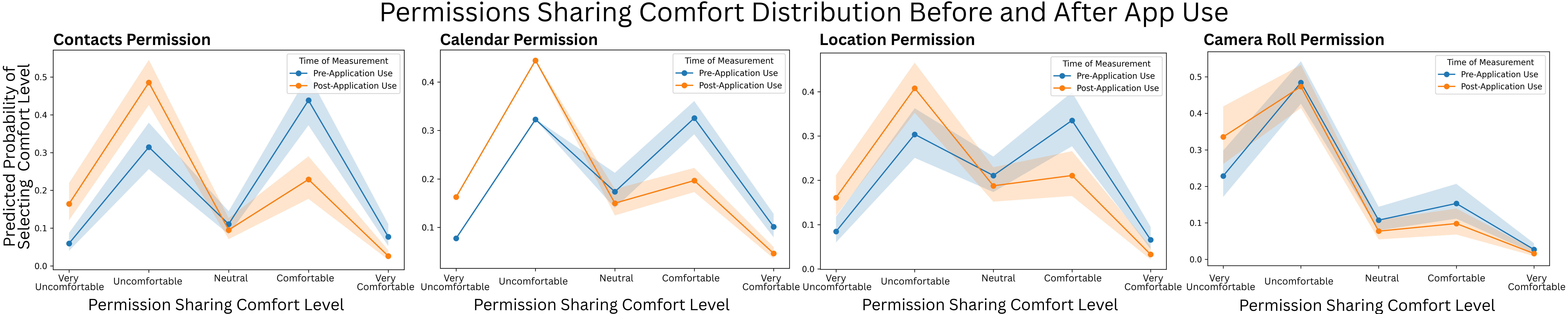}
    \label{fig:eff-time}
\end{figure*}

\textbf{The impact of inference accuracy on user comfort varies across accuracy/comfort conditions.}
When inferences were inaccurate, references to inaccuracy appeared in user justifications at comparable rates across both high and low comfort conditions (47.5\% for high comfort, 52\% for low comfort).
However, when inferences were accurate, users referenced accuracy much more to support increased comfort than to support discomfort (33.6\%, 15.5\%).
This suggests that when inferences are accurate, other features play a larger part in determining discomfort---namely, the inference's sensitivity.

\textbf{When inferences are accurate, comfort is associated with the perceived sensitivity of the data type.}
For accurate inferences, participants with lower comfort frequently reported that inferred data types were sensitive/private (65.3\%). 
For example, in response to an inference about a connection to the substance abuse recovery field, a participant stated,
``this is a stigmatizing topic and I would not want it to be obvious who in my life is in recovery if they are not comfortable with that info being shared.'' 
Similarly, one participant stated they were uncomfortable because ``[their] job involves lots of NDAs and sensitive client information. Having AI drawing data from my personal and work information is very invasive and upsetting.'' 
In contrast, participants with higher comfort ratings often reported that the associated inferences were not sensitive/private (40.7\%). 
For example, in response to an inference regarding their employment sector, one participant stated, ``Most people know this about me and I’m very public with it. Nothing private about it.'' 
We provide more context on specific data types in Appendix~\ref{subsec:inf-category-results}, and also provide more information surrounding their perceived sensitivity.

\textbf{Participants also considered the uniqueness or obviousness of accurate inferences.}
When users reported higher comfort, they frequently mentioned inferences that were perceived to apply to many people; specifically, participants with accurate inferences and higher comfort reported that inferences were not identifying (5.7\%), not specific (3.2\%), and common across populations (3.8\%).
Participants also mentioned that inferences were obvious (6.1\%) or were already available elsewhere (8.5\%), e.g.,, because employment information is already available on LinkedIn. 

\textbf{Some responses addressed potential downstream (positive and negative) consequences of inaccurate inferences.}
Some participants who were uncomfortable with inaccuracy expressed concern with the potential consequences of false information (8.2\%). 
For example, a participant expressed concern about an inference which overestimated their income, stating ``I am so broke. I would not want, for example, the IRS, or some home invader to make such an inference.''
However, some users stated that inaccuracies might be favorable for them (5.9\%); 
for example, a participant who received an inference which overestimated their income said ``I would like to have high income and have people think that.''
Similarly, another participant responded to an incorrect career-related inference by stating ``If it thinks I am more prestigious than I actually am, I’m OK with that.''
The perceived favorability of an inaccurate inference was a driver of participant comfort.

\subsubsection{\textbf{(RQ6): How did the \system app interaction impact participant data sharing?}}
\label{sec:RQ6}

To assess the extent to which
our user study changed the way that
participants thought about privacy,
we analyzed within-participant changes
in reported comfort with
granting data permissions to apps.
We controlled for individual differences
in perceived inference accuracy and
self-reported inference comfort.
Because comfort ratings
were measured on a five-point Likert scale
and collected repeatedly from the same participants,
we modeled the outcome using
a cumulative link mixed-effects model (CLMM)
with a participant-level random intercept
(see Section~\ref{subsubsec:stats}).

Figure \ref{fig:eff-time} displays
the predicted probabilities of data sharing comfort,
with each comfort level plotted separately
for pre-\system-app and post-\system-app conditions. 
The x-axis represents ordered comfort categories,
while the y-axis shows
the model-predicted probability of each response
(similar to a probability density function).
The shaded bands indicate 95\% confidence intervals. 

Across all permission types,
the post-application distributions
place more probability mass on
lower comfort categories and
less on higher comfort categories. 
The most pronounced shift occurs
for the camera roll permission,
and the least pronounced shift occurs
for the contacts permission. 
We note that comfort with sharing for the contacts permission was already quite low to begin with, compared with the other permissions. 
Exposure to the \system application reduced sharing
most amongst participants who found individual inferences uncomfortable. 
Overall,
these results suggest that completing
the \system user study decreased user comfort
with granting data access permissions to apps.
In particular,
comfort decreased the most when
(1) a user thought that \system made accurate inferences, and
(2) those inferences made the user feel uncomfortable.

This trend is supported by our qualitative analysis.
At the end of the user study,
we asked participants to complete
a final, optional open-response question
in which we asked participants 
to report the \textit{ways} in which
they would or would not change
their permission-granting behavior in the future.
52\% of users reported that
their permissions behavior would change,
with 21.5\% users stating that
they better understand permissions data content,
14\% stating they better understand
downstream inference capability, and
16.3\% simply stating that they will be
less willing to share their smartphone data
in the future. 
For example,
one participant stated, 
``Seeing the inferences they could make with the data I allowed, made me definitely more hesitant to allow these permissions. I was surprised they came to the conclusions they did but it was easy to see how they obtained the information. So, it will make me reconsider what permissions.''
Another participate said ``Absolutely. Generally speaking, I already gave limited permissions access to apps prior to this study, but seeing the inferences made by the AI reinforced to me why doing so is important for online privacy.''

Some participants identified reasons for their change,
stating that they found the inference capabilities
of \system to be creepy (2.4\%)
or too successful at identifying
sensitive user characteristics (3.0\%).
Other users identified concerns about
negative impacts from
downstream use of inferred data;
mentioned problems included
surveillance, data breaches,
and third party sharing (9.2\%).
Interestingly,
Thirteen participants stated they would be
\textit{more} comfortable to share data. 
28\% of users
reported that they would not change their behavior;
however, a quarter of this group (7.3\% of the total population)
report that they were already careful
with granting data access to apps. 
A limited set of these participants reported that
they generally did not mind or care about their data privacy (5.4\%), or felt that they had nothing to hide or fear (1.1\%);
however,
the majority of remaining participants
stated they were comfortable with
the insensitivity, inaccuracy, and/or non-identifiability
of the inferences that \system produced about them.

In summary,
both our qualitative and quantitative evaluations
demonstrated that
\textbf{\textit{users are less willing to share
their data with apps if presented with examples
of how that data can be used by LLMs
to create inferences}}.

%% file: paper/6-recommendations-new.tex
We provide recommendations for how smartphone OSes and ML model providers may increase user awareness of downstream privacy impacts of permissions decisions, and mitigate the privacy risks of \system-style inferences.

\paragraph{\textbf{Recommendation \#1: Request-Time Permissions Justifications}}
Modern smartphone OSes require apps to request user consent before accessing sensitive permissions data.
However, as explained in Section~\ref{sec:perms-systems}, Android does not force app developers to provide a \textit{request-time justification} (RTJ) that explains to the user \textit{why} an app needs access to a given permissions data type.
Android's default consent screen simply enumerates the set of permissions the app desires; without an RTJ, users may lack the appropriate context to make an informed consent decision.
While iOS requires apps to provide RTJs, there are minimal requirements on RTJ content (therefore they focus primarily on the \textit{benefits} of data sharing.

Our results suggest that users' permissions decisions are impacted by their understanding of potential downstream data use. 
We therefore recommend that all smartphone OSes should require apps to provide RTJs to users for permissions request pop-ups.
Further, we suggest that these RTJs should focus not just on the \textit{benefits} of sharing, but the \textit{privacy consequences} as well.
For example, OSes could augment third-party RTJs for location data sharing, to state that a third-party will likely be able to infer where they work and live, and infer locations that they visit while the app is running.

\paragraph{\textbf{Recommendation \#2: Privacy Risks Notifications For Newly Created Data}}
A user's phone generates new data over time: contact lists change, new photos are taken, and so on.
We recommend that smartphone OSes run local models over a user's newly-generated local data, flagging new, sensitive data to the user (e.g., a photo of a medical bill), and encouraging them to protect said data where applicable (e.g., moving the sensitive photo to the ``Hidden Photos" album).
While previous work has primarily focused on the data itself, our study demonstrates that user reactions to inferences impact privacy behavior, and OSes should enable transparency into potential downstream data use where possible.

\paragraph{\textbf{Recommendation \#3: Fine-Grained Permissions Data Access Transparency}}
Under current permissions frameworks, it is challenging 
for users to understand when applications are accessing
their permissions data.
Apple recently added an ``App Privacy Report'' feature, which creates a log of each time an app accesses a permissions class. 
However, this logging only tracks access to the permissions class, not specific data items (e.g., to the Camera Roll, not specific photos).
Providing transparency into specific data item access will ease identification of misbehavior of apps conducting \system -like analyses of user data.

\paragraph{\textbf{Recommendation \#4: Update Model Acceptable Use Policies}}
In general,``acceptable use'' policies for popular models do not specifically address the privacy harms that we describe in this paper.
For example, Apple's ``acceptable use'' policy for the on-device Foundation Model framework does not explicitly prohibit models from being leveraged to extract privacy-sensitive user data~\cite{appleFoundationModelsAUP}.
Google's policies for generative AI state that personal data should be not handled ``without legally-required consent''~\cite{googleGenAIaup}.
However, \system-style apps \textit{do} receive consent to access raw personal data.
We are unaware of any Google policy that specifically prohibits the use of Gemini to perform inference-based extraction of private information.
As frontier models and on-device models become more accessible to third-party applications, we recommend that ``acceptable use'' policies explicitly prohibit their use for producing privacy invasive inferences about users.

%% file: paper/8-conclusion.tex
User privacy is increasingly threatened
by the convergence of two trends:
the ubiquity of smartphones,
and the ubiquity of ML models
that can parse multimedia files and unstructured text.
A user's phone
acts a central access point
for diverse types of sensitive data that
are specific to the owner of the device.
Sophisticated ML models,
when granted access to that data,
can infer private aspects of a user's life---aspects that
are financially valuable
to website owners, app developers,
advertisers, and other members
of a vast online ecosystem for
collecting, sharing, and exploiting user data.

In this paper,
we cataloged the types of inferences
that models can make on real-life user data,
and described how users can find those inferences
surprising (or even disturbing).
We demonstrated that many users,
when shown these inferences,
become less willing
to share data with apps in the future.
We also made recommendations for how
smartphone OSes and ML models
should be changed to better respect user privacy.
We hope that this work
sparks a larger conversation
about user privacy
in the midst of widespread ML-driven applications.

%% file: paper/10-ethical-v2.tex
We explore (1) stakeholders and impacts during the research process, (2) impacts of publishing \system‑style findings, (3) mitigations, and (4) justification for the work.
\paragraph{Stakeholders and Process Impact.} \system implicates four primary stakeholder groups as follows. (1) \textit{Study Participants}: Participants installed \system and granted it access to sensitive permissions data (calendar, contacts, photos, location). This data can reveal medical appointments, religious and political activities, financial circumstances, legal issues, and family relationships. During the study, participants face risks of privacy violation and psychological discomfort when confronted with “creepy” or unsettling inferences. (2) \textit{Broader Smartphone Users}: Although we collected data only from participants, the techniques apply to any smartphone user with similar permissions, highlighting privacy risks for a much broader population. (3) \textit{Platform and App Developers}: These stakeholders design and operate the permissions systems and ad‑tech infrastructure that could adopt or resist HARVEST‑style inference pipelines. Our results may influence their technical and business practices, with both beneficial and harmful potential. (4) \textit{Researchers, Policymakers, and Regulators}:
The privacy, security, HCI, and ML research communities, as well as policymakers and regulators, depend on rigorous evidence about inference‑based privacy risks to inform standards, regulation, and oversight.

During the research process, the primary impacts are on participants (via data collection and exposure to inferences) and on platform/app developers (through reputational and regulatory scrutiny). The main impacts on broader users and society arise from the publication and possible replication of \system‑style methods.
\paragraph{Impact of the Research.} \textit{Positive Impacts}. (1) \system makes latent risks visible by documenting what mainstream permissions data can reveal about sensitive attributes (e.g., finances, health, religion, political views, relationships), and how accurate and ``accurate‑ adjacent” inferences are in practice. (2) The app functions as a user education tool: many participants report reduced comfort with permissions sharing after seeing inferences about themselves, particularly when these are accurate and uncomfortable, suggesting that transparency can empower more informed consent. (3) Our results ground concrete recommendations to OS vendors and model providers (e.g., request‑time justifications, model‑level guardrails, clearer acceptable‑use policies) in empirical evidence rather than speculation. (4) Our methodology—jointly analyzing inference accuracy, chain of thought reasoning, and user comfort—provides a framework for future research on LLM‑enabled inference risks.

\textit{Negative Impacts}. (1) By demonstrating that off‑the‑shelf, open‑ source models with limited compute and data can already yield detailed inferences, we highlight an attack surface that other entities or data brokers could exploit, further intensifying profiling and surveillance. (2) Our analysis of “accurate‑adjacent” and incorrect assumptions also shows that even erroneous inferences can be highly valuable for targeting, potentially normalizing systems that treat misclassification as acceptable collateral. (3) Lowering the barrier for non‑expert adversaries: describing a working pipeline risks enabling less experienced actors to replicate \system‑style inference with minimal additional effort. (4) Some participants experienced discomfort or distress upon seeing sensitive, detailed inferences and the reasoning behind them.
\paragraph{Mitigations.} \textit{Methodological Mitigations (Implemented)}. (1) We limited the quantity and kinds of data collected (e.g., restricting the number of photos) and focused on a defined subset of permissions rather than broad device access. (2) Consent materials described the categories of data accessed and the kinds of inferences that might be generated (including potentially sensitive and uncomfortable ones), enabling informed participation. (3) All permissions data and model outputs remained within a controlled research environment, accessible only to authorized researchers under institutional protocols. Data were not shared with third parties, commercial systems, or data brokers, and did not leave our contained system. (4) We report the study design and analysis at a level sufficient for scientific understanding but do not release a full \system codebase, detailed platform implementation, or a turnkey recipe for deployment. Our goal is to study and evidence the risk, not to provide a ready‑to‑use targeting toolkit.

\textit{Recommended Future Deployment Measures}. For real‑world systems, we recommend: (1) OS‑level tools that let users preview inferences before granting permissions, ideally on‑device; (2) consent mechanisms that separately address raw data access versus inference of sensitive attributes; (3) model‑level guardrails that restrict undeclared inference of sensitive traits from both overt and latent signals; and (4) clearer, enforceable acceptable‑use and documentation standards that explicitly cover inference‑based privacy harms.
\paragraph{Justification for Research.} Permissions‑based data collection and LLM deployment are already widespread; without empirical study, users and regulators lack evidence about what current systems can infer. Our design choices intentionally avoid creating a high‑ performance, deployable targeting system. With the study, we show that exposure to these inferences can improve user understanding and reduce permissiveness, suggesting a path toward user‑empowering transparency tools. Providing concrete technical and policy recommendations aid in the mitigation of \system‑style privacy harms.

Our university’s Institutional Review Board reviewed and approved this study, including a Data Safety Review of encryption, access restrictions, and data retention practices.

%% file: paper/11-osf.tex
In this work, we carefully considered the ethical implications of releasing certain research artifacts, particularly system prompts and raw participant data. 
While transparency and reproducibility are central values of open science, unrestricted release of these materials could facilitate the development or deployment of systems capable of generating invasive or privacy-compromising inferences about individuals. 
To mitigate this risk, we chose not to publicly release the full set of prompts used in our study.

Similarly, due to the sensitivity of the data collected, we do not share extended, non-aggregated participant data in order to safeguard user privacy and thwart potential misuse. 
Instead, we provide selectively redacted and aggregated materials in the appendix to support transparency while minimizing the risk of re-identification or harm.

These decisions reflect an ongoing effort to balance openness with responsibility. We recognize the importance of enabling scientific scrutiny and replication, but we also seek to reduce the likelihood that bad actors could repurpose our artifacts “off the shelf” to build systems that undermine individual autonomy or privacy, or cause any harm to participants. 
We are therefore continuing to evaluate which subsets of artifacts can be safely released without amplifying these risks.

%% file: paper/9-appendix.tex
\section{Social Media App Permission Requests}
\label{companies-collect}
A study by PEW Research identifies that the five most popular social media applications are Youtube, Facebook, Instagram, TikTok, and WhatsApp~\cite{pew-socialmedia}. 
We provide the data types collected by these applications below. 

\begin{table}[h!]
\centering
\begin{tabular}{c c c c c c}
\toprule
Data Type & YT & FB & IG & TT & WA \\
\midrule
\makecell{Device and app history}  &  & X &  &  & X \\
\hline 
\makecell{Identity}  & X & X & X &  & X \\
\hline
\makecell{Calendar}  &  & X & X &   &  \\
\hline 
\makecell{Contacts}  & X  & X  & X & X & X \\
\hline
\makecell{Location}  & X & X & X & X & X \\
\hline 
\makecell{SMS}  &  &  &  &  & X \\
\hline 
\makecell{Phone}  & X  & X & X &   & X \\
\hline 
\makecell{Photos/Media/Files}  & X & X & X & X & X \\
\hline 
\makecell{Storage}  &  X & X & X & X & X \\
\hline 
\makecell{Camera}  & X & X & X & X & X \\
\hline 
\makecell{Microphone} & X & X & X & X & X \\
\hline 
\makecell{Wi-Fi connection information} & X & X & X & X & X \\
\hline 
\makecell{Device ID \& call information} & X & X & X &  & X \\
\bottomrule
\end{tabular}
\caption{Data types collected by popular social media applications, as listed on the Google Play Store as of early 2026. YT=YouTube, FB=Facebook, IG=Instagram, TT=TikTok, WA=WhatsApp.}
\end{table}

\noindent In our study, we investigate:
\begin{itemize}
    \item Calendar (collected by Facebook, Instagram)
    \item Contacts (collected by Youtube, Facebook, Instagram, TikTok, WhatsApp)
    \item Location (collected by Youtube, Facebook, Instagram, TikTok, WhatsApp)
    \item Photos  (collected by Youtube, Facebook, Instagram, TikTok, WhatsApp)
\end{itemize}

\section{Sample Inference by Type}
\label{example-infs}

We provide a sample of accurate (unless stated otherwise) inferences provided by participants, redacted to maintain participant privacy. 

\subsection{Mental Health} 

\textbf{Inference: }You have a history of struggling with attention deficit hyperactivity disorder (ADHD) and are currently taking Concerta to manage your symptoms.
\\
\textbf{Reasoning: }The calendar data includes a note about taking Concerta and experiencing side effects, which suggests the user is taking medication for ADHD. Additionally, the calendar event titled `Happy 2 Months on Concerta' suggests they are celebrating a milestone in their treatment.
\\
\\
\textbf{Inference: }You or someone close to you has experienced trauma or stress related to memory loss and is seeking information on C-PTSD.
\\
\textbf{Reasoning: }The image shows a post from @[PTSD related social media account], which suggests that the user is interested in learning about C-PTSD and its effects on memory, indicating a personal connection to the topic.
\\
\\
\textbf{Inference: }You prioritize your mental health and well-being, as evidenced by your history of therapy sessions.
\\
\textbf{Reasoning: }The calendar data shows a history of therapy sessions.
\\
\\
\textbf{Inference: }You or someone close to you has struggled with OCD and is seeking support for mental health issues.
\\
\textbf{Reasoning: }The image shows a conversation about OCD and therapy, indicating your interest in mental health support.
\\
\\
\textbf{Inference: }You are seeking ongoing mental health treatment, including therapy and psychiatric care.
\\
\textbf{Reasoning: }The high frequency of `Counseling appointment' and `Psychiatrist appointment' events in the user's calendar data suggests that they are seeking ongoing mental health treatment.
\\
\\
\textbf{Inference: }You have attended counseling sessions online via Zoom, specifically with [Redacted: Therapist Name] on [Redacted: MM-DD-YYYY], and [Redacted: Therapist Name] on [Redacted: MM-DD-YYYY].
\\
\textbf{Reasoning: }The calendar data shows counseling appointments with [Redacted: Name] on [Redacted: MM-DD-YYYY], and [Redacted: Name] on [Redacted: MM-DD-YYYY], via Zoom, supporting the hypothesis.
\\
\\

\subsection{Medical Conditions}
\textbf{Inference: }You have a family member who has been dealing with a serious medical condition, specifically cancer, and has undergone various medical tests and appointments.
\\
\textbf{Reasoning: }The calendar data shows multiple appointments with doctors, including oncology and cardiology specialists, and medical tests such as bloodwork and a tilt table test, suggesting that a family member has been dealing with a serious medical condition.\\
\\
\textbf{Inference: }You or someone close to you is taking Ozempic, a medication commonly used for type 2 diabetes or weight loss, on a regular basis.
\\
\textbf{Reasoning: }The calendar data lists a repeated `[Redacted: Name] ozempic' event, which suggests a consistent medication schedule (calendar source).
\\
\\
\textbf{Inference: }You or someone close to you has a medical condition related to rheumatoid arthritis.
\\
\textbf{Reasoning: }The presence of `Rheumatoid Center' in the organization names suggests that the user or someone close to them has a medical condition related to rheumatoid arthritis.
\\
\\
\textbf{Inference: }You have a rare medical condition and have sought connections with others who share similar experiences.
\\
\textbf{Reasoning: }The presence of [Redacted: Company Name] in the organization names suggests... a rare medical condition, and the user's involvement with this organization indicates a desire to connect with others who share similar experiences.
\\
\\

\subsection{About Children}
\textbf{Inference: }You or someone close to you has a child with special needs. 
\\
\textbf{Reasoning: }The presence of `Speech therapist', `[Redacted: Clinic Name]', and `[Redacted: Clinic Name] - Physical therapist' in the organization names suggests that the user or someone close to them has a child with special needs.
\\
\\
\textbf{Inference: }You or someone close to you has a child with a genetic condition and has sought medical attention for this condition.
\\
\textbf{Reasoning: }The presence of `Pediatric Medical Genetics' as an organization name in the user's contacts data, combined with the user's location in South [Redacted: State], suggests a connection to a child with a genetic condition and access to medical attention.
\\
\\
\textbf{Inference: }You are a parent or guardian of a child who has had oral surgery.
\\
\textbf{Reasoning: }The event `[Redacted: Name] oral surgery' on [Redacted: MM-DD-YYYY], indicates that the user or someone close to them has had oral surgery, and the presence of other events related to children (e.g., `[NAME] first birthday!') suggests that the user may be a parent or guardian. (calendar data).
\\
\\
\textbf{Inference: }You or someone close to you has a child with a genetic condition and has sought medical attention for this condition.
\\
\textbf{Reasoning: }The presence of `Pediatric Medical Genetics' as an organization name in the user's contacts data, combined with the user's location in [Redacted: Region of State], suggests a connection to a child with a genetic condition and access to medical attention.
\\
\\

\subsection{Intimate Content}
\textbf{Inference: } You have a strong interest in fashion or beauty and have likely attended events or followed influencers like Sophie Rain in the past.
\\
\textbf{Reasoning: }The combination of Sophie Rain's fashion and beauty-focused posts and the user's interest in her content suggests a strong interest in these topics, as evidenced by the user's camera roll.
\footnote{This inference was marked as ``Somewhat inaccurate'' by the participant. We see that the model was missing the context that the user was a 25-34 year old male.}
\\
\\
\textbf{Inference: }You have received a romantic text message from someone who is interested in you.
\\
\textbf{Reasoning: }The text message contains flirtatious language and suggests a romantic interest in the user. The message also includes a phone number, which could be used to contact the user directly. (Text message source)
\\
\\
\textbf{Inference: }You are interested in dating and are using a dating app to connect with others.
\\
\textbf{Reasoning: }The dating app profile picture and interface suggest that you are interested in dating and are using the app to connect with others. The app's features and settings are also visible on the screen.

\subsection{Reproductive Health}
\textbf{Inference: }You are a woman who has experienced issues with birth control and is seeking alternative methods. 
\\
\textbf{Reasoning: }The user's comment about having issues with birth control and seeking alternative methods suggests that they are a woman who has experienced difficulties with traditional birth control methods. The user's mention of a holistic doctor and their interest in finding a more natural approach to health and wellness also supports this hypothesis.
\\
\\
\textbf{Inference (as summarized by user): } my sister in law is pregnant
\\
\textbf{Reasoning: }The calendar data lists an event titled `[Redacted: Name] Ultrasound' on [Redacted: MM-DD-YYYY], which suggests a medical appointment, possibly related to pregnancy.
\\
\\
\textbf{Inference: }You are interested in prenatal care or parenting resources, which may indicate that you are preparing for a new addition to your family.
\\
\textbf{Reasoning: }The image of a website related to prenatal care or parenting resources in the user's camera roll and the ultrasound image suggest that the user is interested in prenatal care or parenting resources, which may indicate that they are preparing for a new addition to their family.

\subsection{Gambling}
\textbf{Inference: }You have placed a bet on a sports game, specifically the [Redacted: Sports Team], indicating an interest in sports betting.
\\
\textbf{Reasoning: }The sports betting app and profile picture indicate that you have placed a bet on a sports game and are interested in sports betting.
\\
\\
\textbf{Inference: }You have placed a bet on a recent or upcoming NFL game.
\\
\textbf{Reasoning: }The image shows a screenshot of a sports betting app, with a list of recent bets and their outcomes, which suggests you have placed a bet on a recent NFL game.

\subsection{Financial Status}
\textbf{Inference: }You are trying to access a cryptocurrency wallet.
\\
\textbf{Reasoning: }The image shows a screenshot of a mobile device with a cryptocurrency wallet app open. The app is displaying a private key, which is a unique code used to access the wallet. (camera roll source)
\\
\\
\textbf{Inference: }You have sent a significant amount of money to someone named [Redacted: Name], possibly for business or personal reasons.
\\
\textbf{Reasoning: }The image shows a screenshot of a mobile payment app, with your name and the recipient's name ([Redacted: Name]) visible, and the amount of money being sent is also visible.
\\
\\
\textbf{Inference: }You have applied for nutrition benefits, indicating that you or someone in your household may be experiencing financial difficulties or food insecurity.
\\
\textbf{Reasoning: }The calendar data includes an entry for applying for nutrition benefits on [Redacted: State Health Benefits Website], suggesting that the user or someone in their household may be experiencing financial difficulties or food insecurity.
\\
\\
\textbf{Inference: }You or someone close to you is applying for Medicare.
\\
\textbf{Reasoning: }The image shows a screenshot of the Medicare website, which suggests that you or someone close to you is applying for Medicare.
\\
\\
\textbf{Inference: }You or someone close to you has a financial or loan-related issue, possibly with a mortgage or other large debt.
\\
\textbf{Reasoning: }The user's contacts data shows contacts from financial or loan-related services, such as ameriprofunding.com and usadiscounters.net, suggesting a connection to financial or loan-related issues.
\\
\\
\textbf{Inference: }You have made purchases using installment plans, such as Zip and QuadPay, indicating a possible reliance on credit or limited financial flexibility.
\\
\textbf{Reasoning: }The calendar data shows multiple events related to installment payments, such as `Zip Installment due for order [Redacted: Order Number]', which suggests a reliance on credit or limited financial flexibility.
\\
\\
\textbf{Inference: }You have a car insurance policy with [Redacted: Insurance Company] that is effective from [Redacted: MM-DD-YYYY], to [Redacted: MM-DD-YYYY], and have insured a Chrysler Fiat 500 with a policy number of [Redacted: Policy Number].
\\
\textbf{Reasoning: }The image shows a [Redacted: Car Insurance Company] card with the user's name policy information, and car information...
\\
\\
\textbf{Inference: }You have a past due balance on your [Redacted: Electricity Company] [Redacted: State] account.
\\
\textbf{Reasoning: }The image shows a past due balance on the [Redacted: Electricity Company] [Redacted: State] account, and the user's browser history shows that they have been searching for information about their energy bill and payment history.
\\
\\

\subsection{Legal Issues}
\textbf{Inference: }You are a resident of [Redacted: Location] Residential Corrections.
\\
\textbf{Reasoning: }The image shows an ID card with the name `[Redacted: Name]' and the number `\#[Redacted: ID Number]'. The card also mentions '[Redacted: Location] Residential Corrections'...This suggests that you are a resident of the facility.
\\
\\
\textbf{Inference: }You are likely involved in a court case or legal proceedings.
\\
\textbf{Reasoning: }The event `[Redacted: Region] court at 1 pm in [Redacted: Department]' on [Redacted: MM-DD-YYYY], strongly suggests that the user is involved in a court case or legal proceedings. (calendar source)
\\
\\
\textbf{Inference: }You or someone close to you has dealt with immigration issues and may have sought legal advice.
\\
\textbf{Reasoning: }The presence of `[Translated: Immigration Lawyers]' as an organization name in the user's contacts data suggests a connection to immigration issues, and the user may have sought legal advice or services related to immigration.
\subsection{Politics}
\textbf{Inference: }You are likely involved in local politics, possibly as a precinct chair or volunteer, and have a strong connection to the Democratic party.
\\
\textbf{Reasoning: }The user's contacts include a label `[Redacted: ID] Precinct Chair' and an organization name `Democratic Poll Greeting', which suggests a strong connection to the Democratic party. The user's location is in a neighborhood with a moderate median household income, according to data from the US Census Bureau.
\\
\\
\textbf{Inference: }You are likely a conservative or have connections to conservative organizations.
\\
\textbf{Reasoning: }The presence of `[Redacted: State-Level Pro-Life/Family Organization]' in the organization names suggests a connection to conservative organizations. (contacts data)
\subsection{Sexual Orientation}

\textbf{Inference: }You or someone close to you identifies as part of the LGBTQ+ community.
\\
\textbf{Reasoning: }The presence of Grindr in the user's contacts data, a popular LGBTQ+ dating app, supports the hypothesis that the user or someone close to them identifies as part of the LGBTQ+ community.

\subsection{Religion}
\textbf{Inference: }You or someone close to you has a connection to the Jewish community.
\\
\textbf{Reasoning: }The user has multiple contacts with Jewish-sounding names and organizations, such as `[Redacted: Template Name]' and `Jewish Federation of [Redacted: City]'.
\\
\\
\textbf{Inference: }You or someone close to you is involved in a community or organization related to Unitarian Universalism.
\\
\textbf{Reasoning: }The user has multiple events related to [Redacted: Acronym]([Redacted: Region] Unitarian Universalists), including `[Redacted: Event Name] at [Redacted: Acronym]' and `[Redacted: Acronym] - [Redacted: Event Name]', which suggests involvement in this community (calendar source).
\\
\\
\textbf{Inference: }You are a member of the Church of Jesus Christ of Latter-day Saints.
\\
\textbf{Reasoning: }The location data shows a high concentration of members of the Church of Jesus Christ of Latter-day Saints in the area, which may indicate the user is a member of the church (location source).
\subsection{Demographics}
\textbf{Inference: }You have a connection to the [Redacted: Tribe Name] Indian People and may have a history of financial struggles or debt.
\\
\textbf{Reasoning: }The presence of multiple contacts with `[Redacted: Tribe Name] Indian People' suggests the user has a connection to this organization, and the presence of contacts with `Debt' and `Financial' labels suggests the user or someone close to them has a history of financial struggles or debt.
\\
\\
\textbf{Inference: }You have an interest in African American culture.
\\
\textbf{Reasoning: }The product's name and description, `Shoe String King Square Afro Pick with Black Fist', imply a connection to African American culture.
\\
\\
\textbf{Inference: }You are likely of Hispanic or Latino descent.
\\
\textbf{Reasoning: }The location data indicates that the user is in an area with a high Hispanic population, which could suggest that the user is of Hispanic or Latino descent.
\\
\\

\subsection{Interests}
\textbf{Inference: }You are likely a gun owner or have an interest in firearms...
\\
\textbf{Reasoning: }The numerous contacts related to firearms and gun enthusiast groups (e.g., `[Redacted: Shooting Range], `[Redacted: Gun Club]') indicate that the user is likely a gun owner or has an interest in firearms.
\\
\\
\textbf{Inference: }You or someone close to you has a strong interest in social justice and activism, particularly in the area of racial equality.
\\
\textbf{Reasoning: }The user's calendar data shows a past event titled `[Redacted: Black Studies Dept Webinar Focused On Black Male Studies]', which suggests a strong interest in social justice and activism, particularly in the area of racial equality...
\\
\\
\textbf{Inference: }You have a tattoo on your left shoulder.
\\
\textbf{Reasoning: }The image shows a tattoo on your left shoulder, which is visible in the top right corner of the image (image source).
\\
\\

\section{App Screenshots}
\label{app-screenshots}

\begin{figure*}[t]
    \centering
    \setlength{\tabcolsep}{6pt}
    \renewcommand{\arraystretch}{1}

    \begin{tabular}{ccc}
        \includegraphics[width=0.25\textwidth]{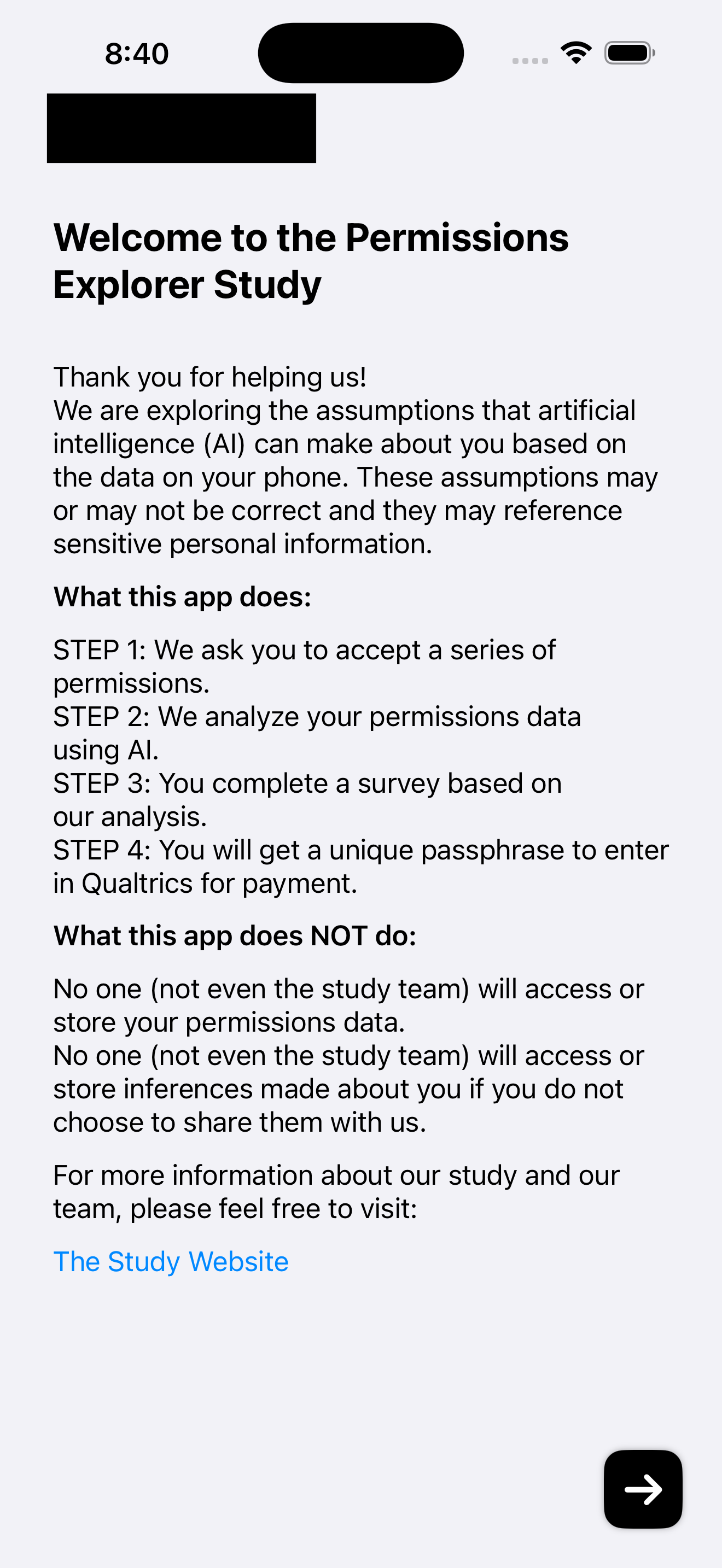} &
        \includegraphics[width=0.25\textwidth]{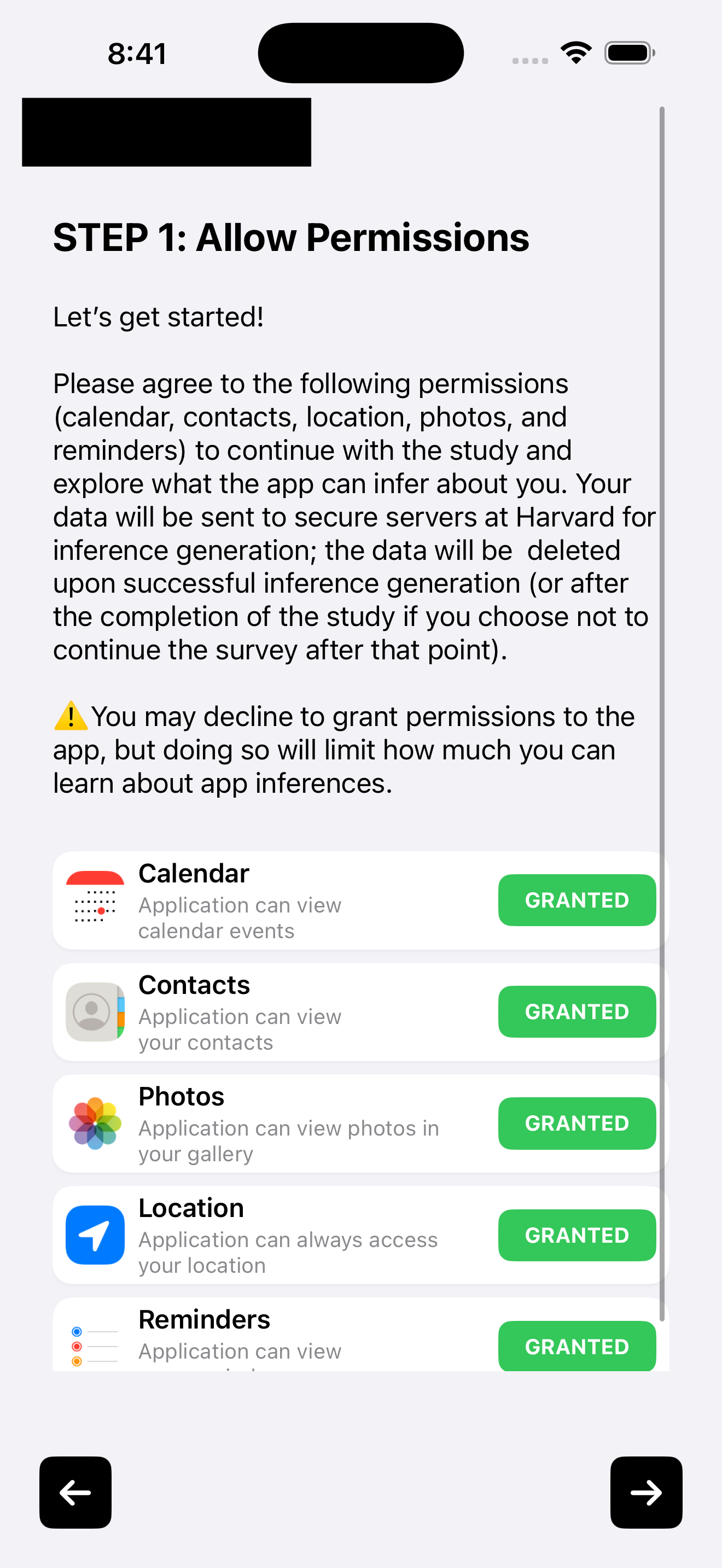} &
        \includegraphics[width=0.25\textwidth]{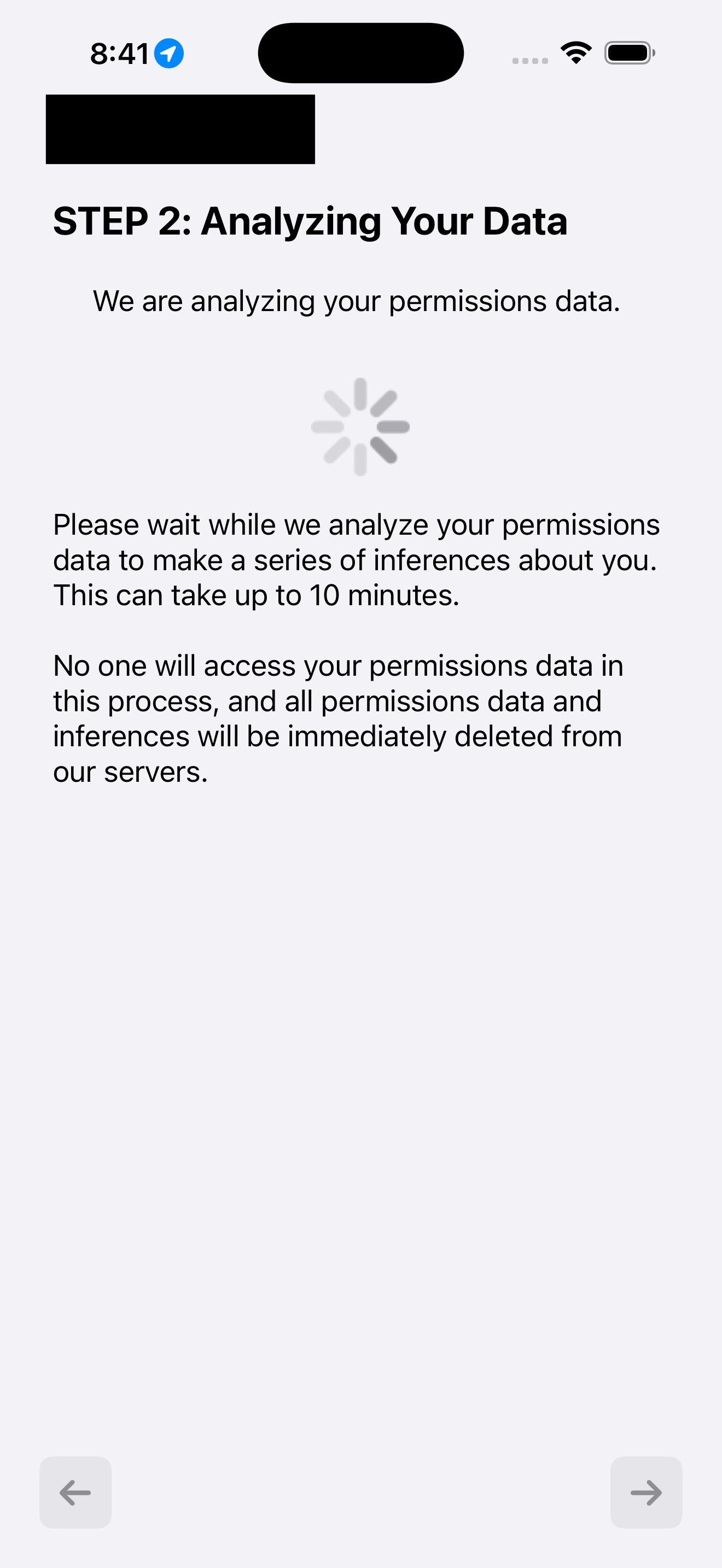} \\

        \includegraphics[width=0.25\textwidth]{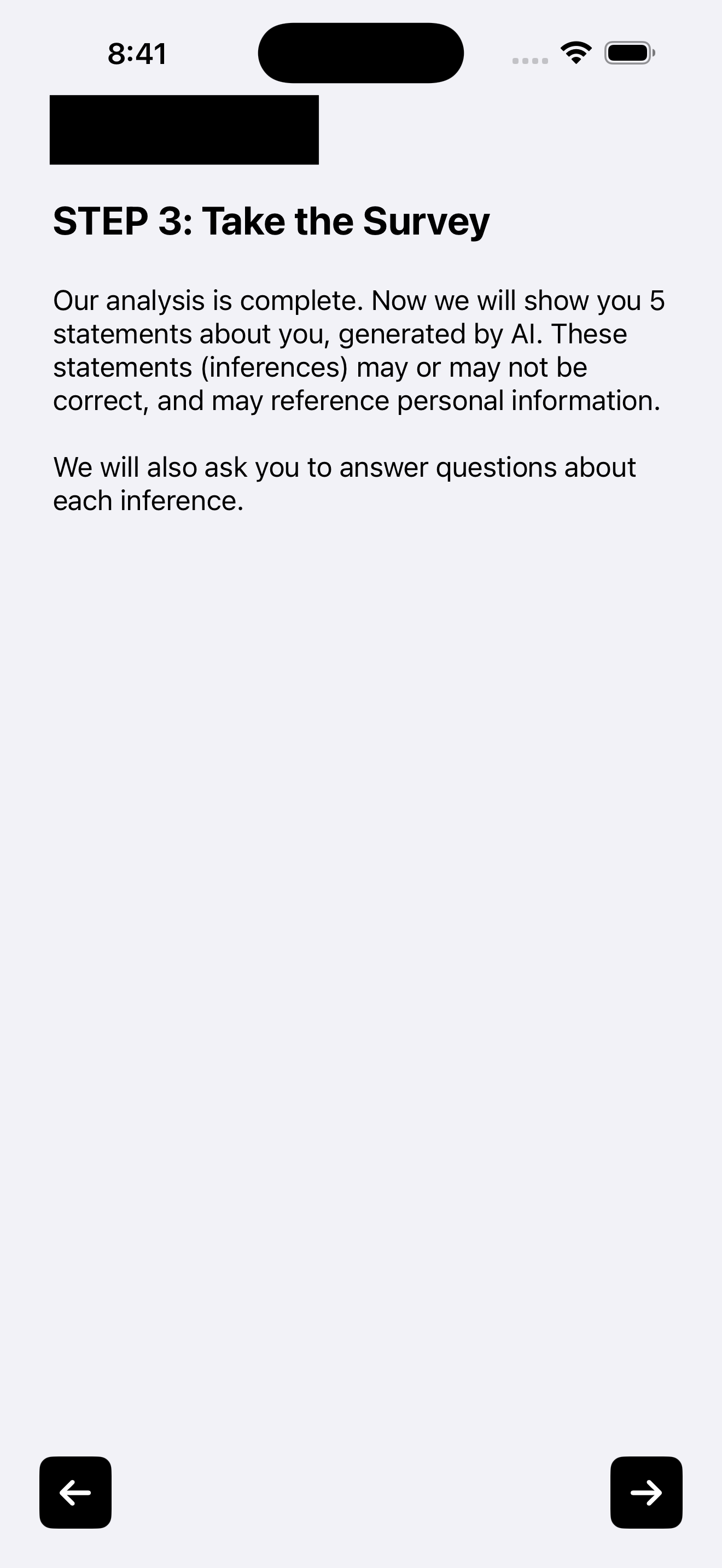} &
        \includegraphics[width=0.25\textwidth]{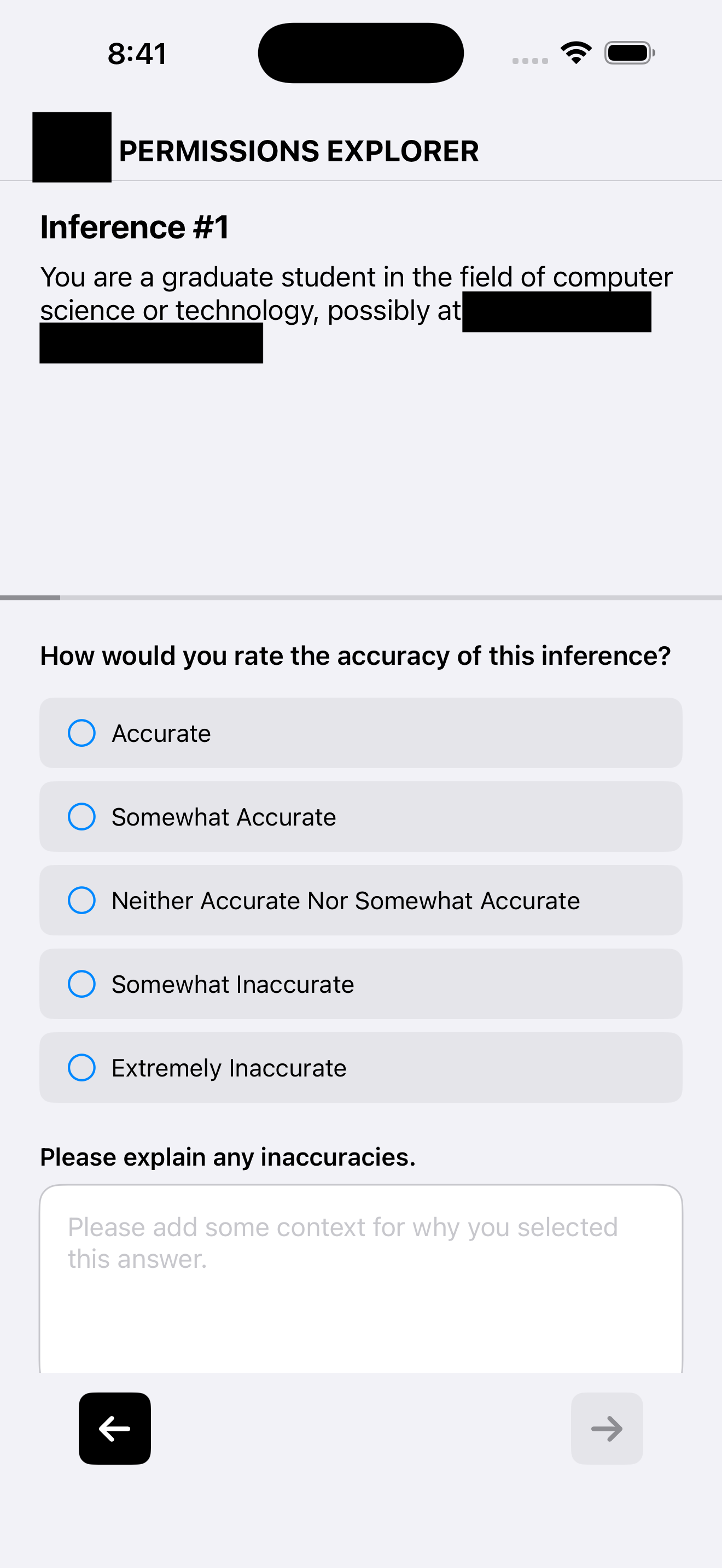} &
        \includegraphics[width=0.25\textwidth]{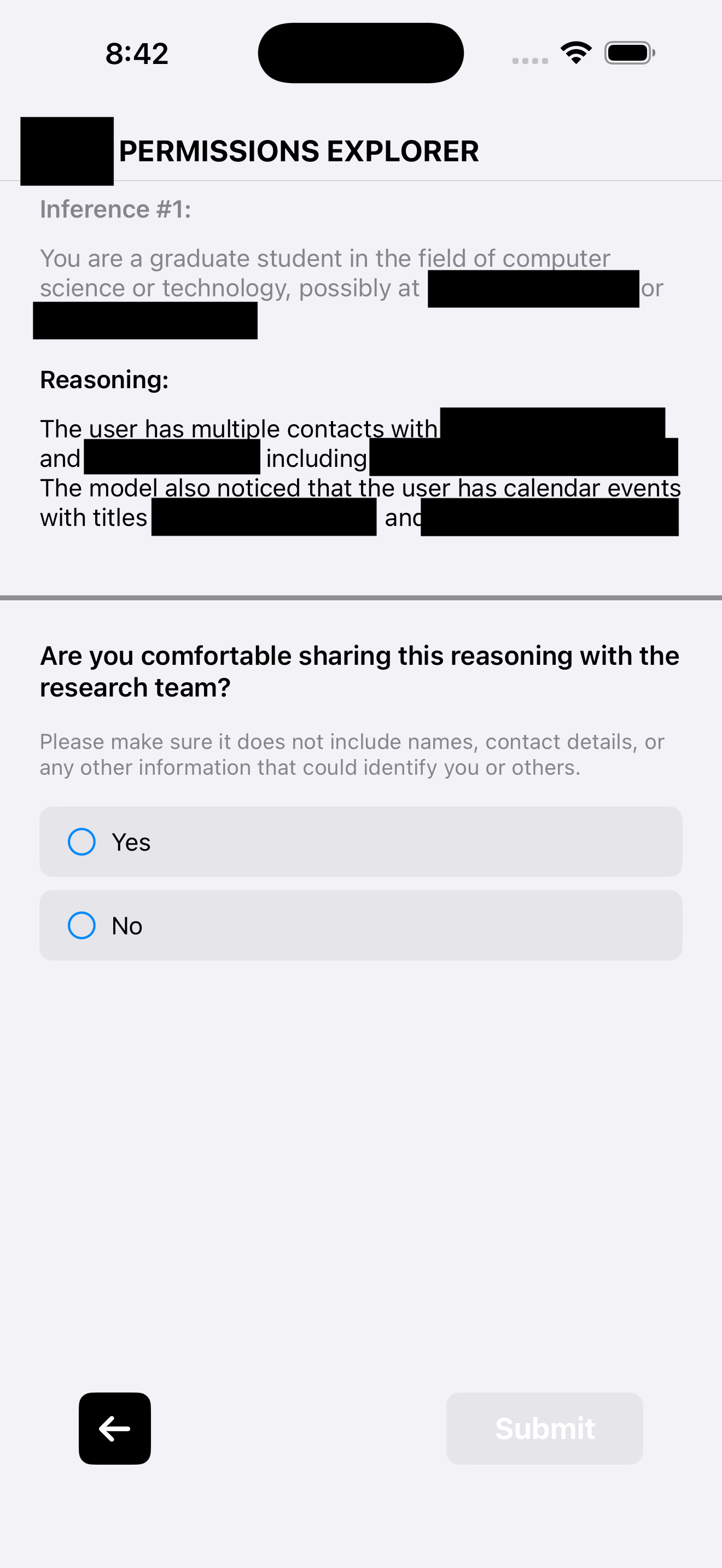}
    \end{tabular}

    \caption{Screenshots of the iOS application interface.\protect\footnotemark[1]}
    \label{fig:screenshots-android}
\end{figure*}

\footnotetext[1]{Identifying information redacted to maintain anonymity of the research team.}

\begin{figure*}[t]
    \centering
    \setlength{\tabcolsep}{6pt}
    \renewcommand{\arraystretch}{1}

    \begin{tabular}{ccc}
        \includegraphics[width=0.25\textwidth]{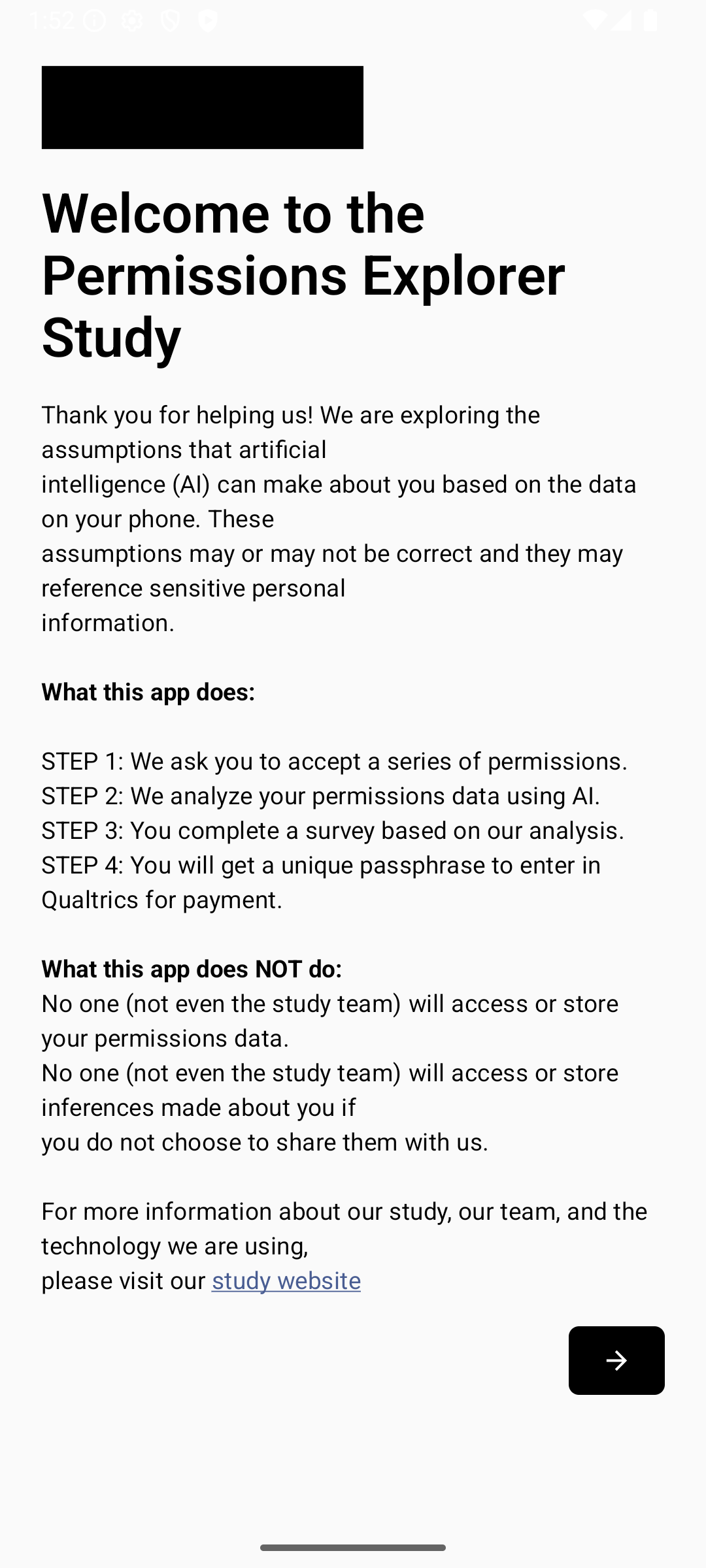} &
        \includegraphics[width=0.25\textwidth]{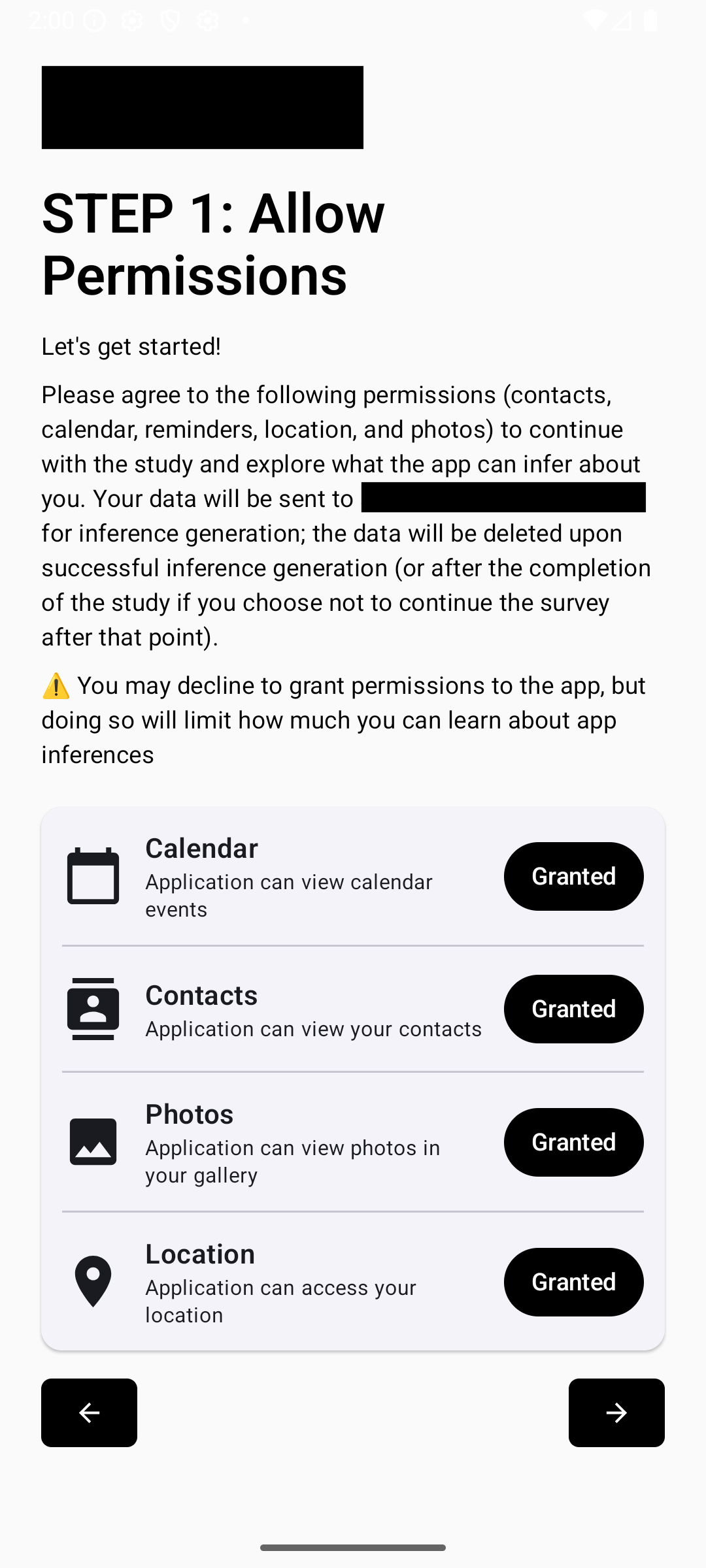} &
        \includegraphics[width=0.25\textwidth]{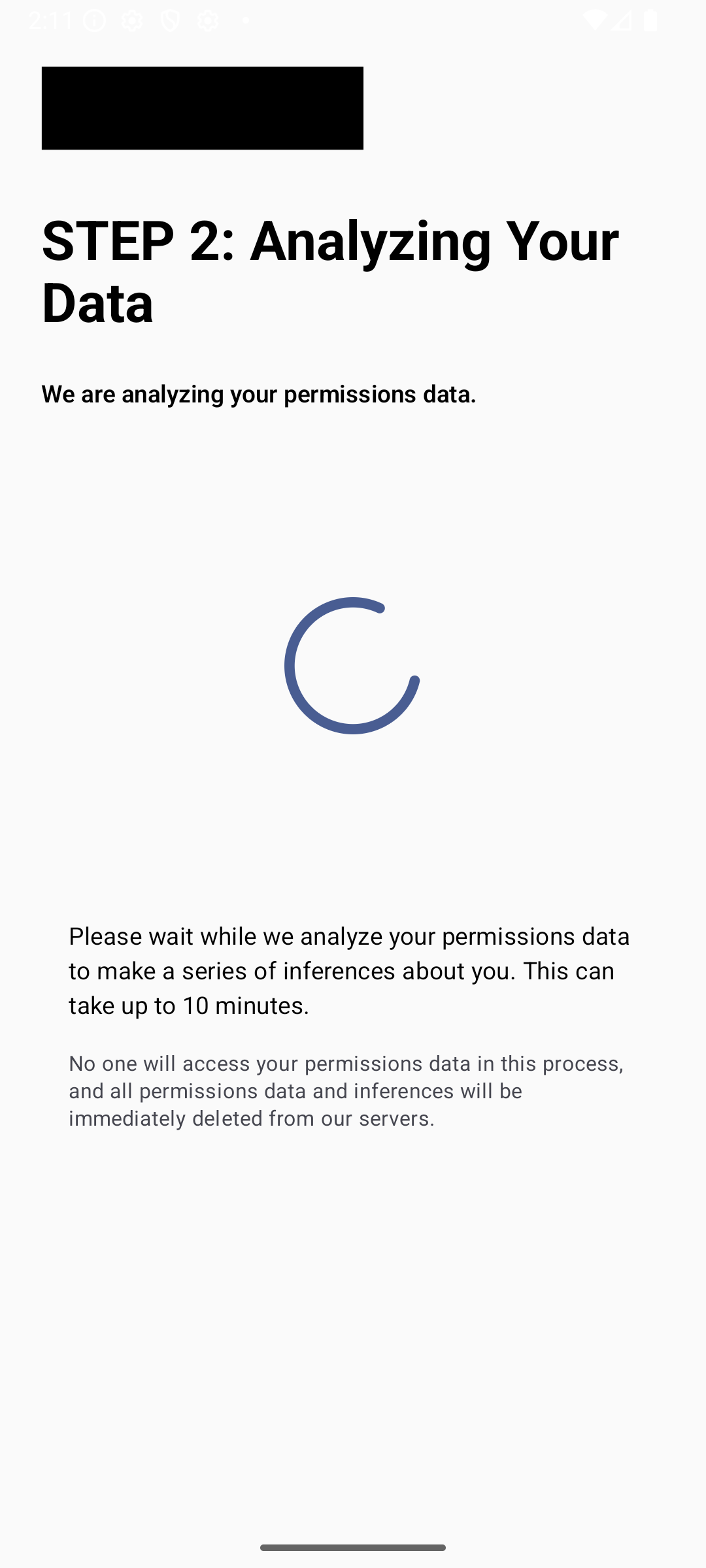} \\

        \includegraphics[width=0.25\textwidth]{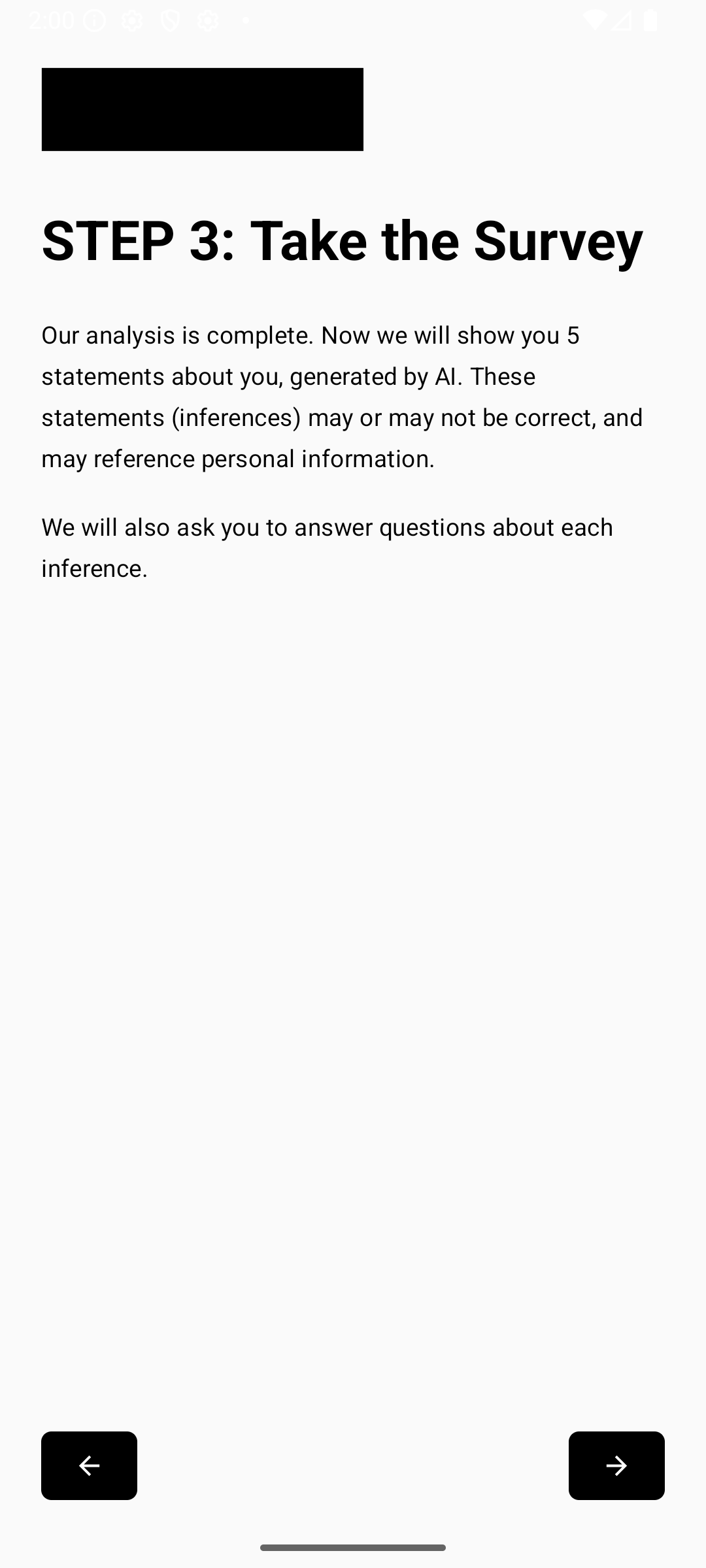} &
        \includegraphics[width=0.25\textwidth]{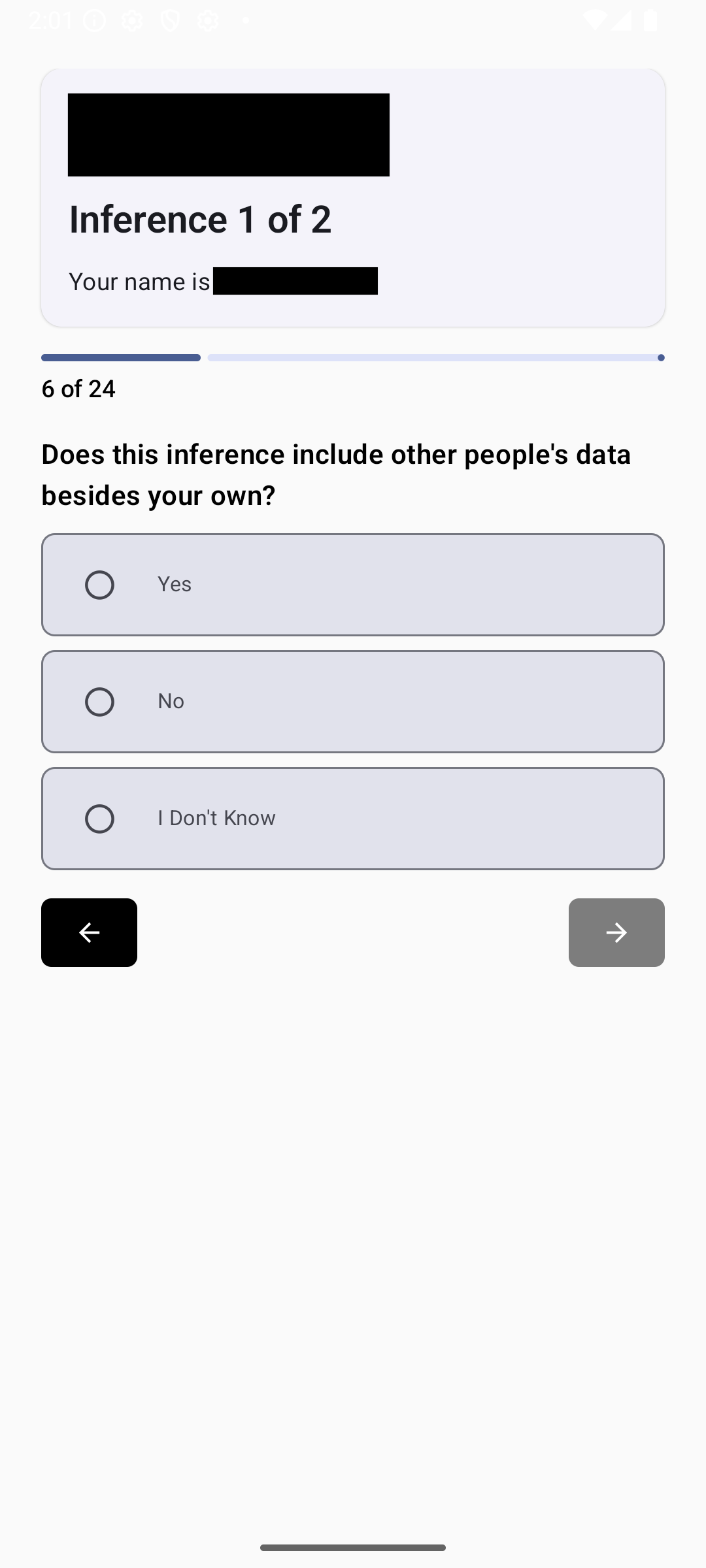} &
        \includegraphics[width=0.25\textwidth]{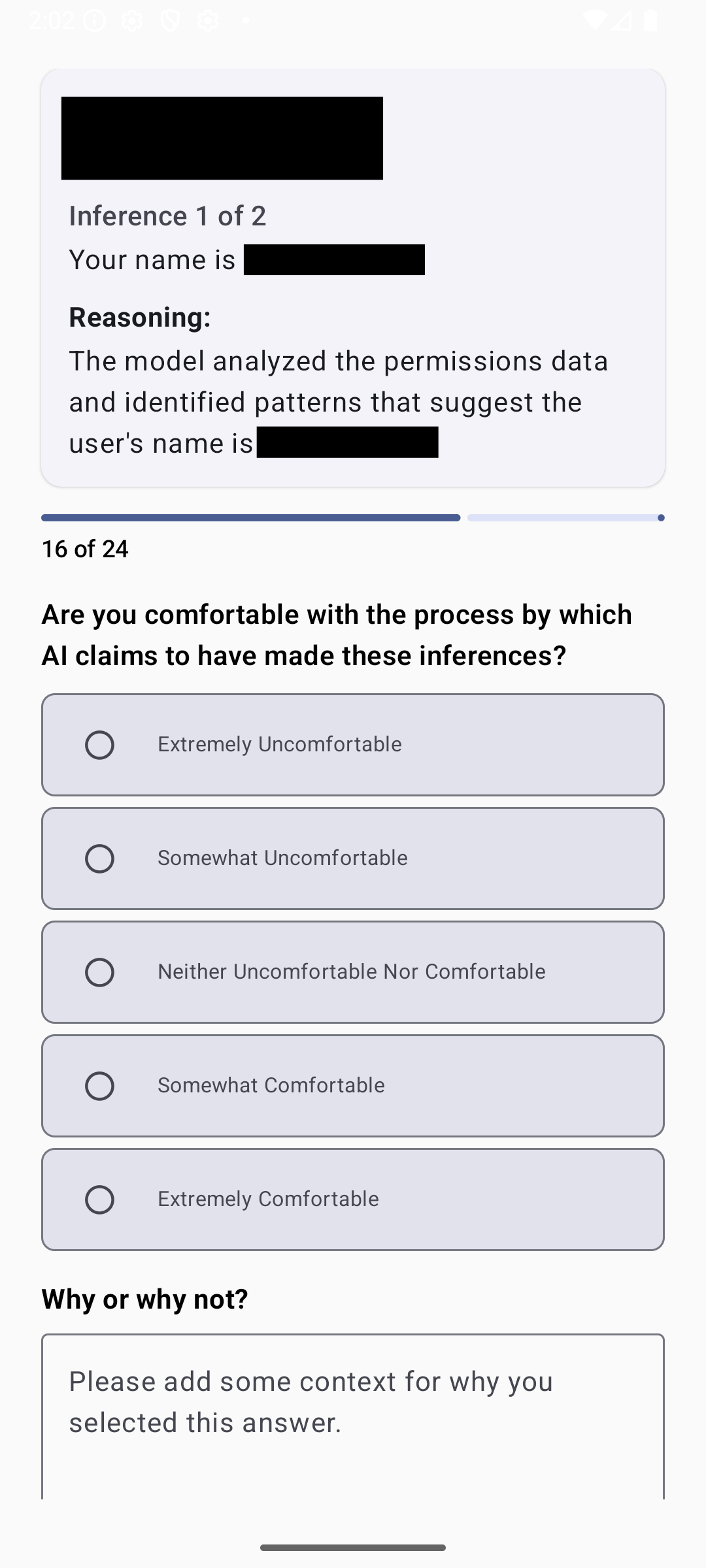}
    \end{tabular}

    \caption{Screenshots of the Android application interface.\protect\footnotemark[1]}
    \label{fig:screenshots-android}
\end{figure*}

\clearpage

\section{In-App Survey Questions}
\label{survey-questions}

The following survey questions were presented to participants regarding AI inferences:

\subsection{Inference Questions}
\begin{description}
    \item[Q0] How would you rate the accuracy of this inference?\\
    Options: 
    \begin{itemize}
        \item Accurate
        \item Somewhat Accurate
        \item Neither Accurate Nor Somewhat Accurate
        \item Somewhat Inaccurate
        \item Extremely Inaccurate
    \end{itemize}

    \item[Q1] Are you comfortable with this inference made by AI?\\
    Options:
    \begin{itemize}
        \item Extremely Uncomfortable
        \item Somewhat Uncomfortable
        \item Neither Uncomfortable Nor Comfortable
        \item Somewhat Comfortable
        \item Extremely Comfortable
    \end{itemize}

    \item[Q2] Would you be surprised if a person who looked at your data could make this inference?\\
    Options:
    \begin{itemize}
        \item Extremely Surprised
        \item Somewhat Surprised
        \item Neither Surprised Nor Unsurprised
        \item Somewhat Not Surprised
        \item Extremely Not Surprised
    \end{itemize}

    \item[Q3] Are you surprised an app can use code to make this inference?\\
    Options:
    \begin{itemize}
        \item Extremely Surprised
        \item Somewhat Surprised
        \item Neither Surprised Nor Unsurprised
        \item Somewhat Not Surprised
        \item Extremely Not Surprised
    \end{itemize}

    \item[Q4] Which of these categories describe this inference?\\
    Options:
    \begin{itemize}
        \item Identity (e.g., race, gender, age)
        \item Activities / interests
        \item Past or current location
        \item Health
        \item Relationships
        \item Education / career
        \item Finances
        \item Beliefs / politics
        \item Personally Identifiable Information (e.g., email, phone, DOB, name, SSN)
    \end{itemize}

    \item[Q5] Does this inference include other people's data besides your own?\\
    Options:
    \begin{itemize}
        \item Yes
        \item No
        \item I Don't Know
    \end{itemize}

    \item[Q6] Are you comfortable sharing this inference text with the research team?\\
    Options:
    \begin{itemize}
        \item Yes
        \item No
    \end{itemize}
\end{description}

\subsection{Chain of Thought Questions}

The survey questions presented to participants regarding model reasoning are listed below:

\begin{description}
    \item[Q0] How would you rate the accuracy of this inference?\\
    Options:
    \begin{itemize}
        \item Accurate
        \item Somewhat Accurate
        \item Neither Accurate Nor Somewhat Accurate
        \item Somewhat Inaccurate
        \item Extremely Inaccurate
    \end{itemize}

    \item[Q1] Are you comfortable with this inference made by AI?\\
    Options:
    \begin{itemize}
        \item Extremely Uncomfortable
        \item Somewhat Uncomfortable
        \item Neither Uncomfortable Nor Comfortable
        \item Somewhat Comfortable
        \item Extremely Comfortable
    \end{itemize}

    \item[Q2] Would you be surprised if a person who looked at your data could make this inference?\\
    Options:
    \begin{itemize}
        \item Extremely Surprised
        \item Somewhat Surprised
        \item Neither Surprised Nor Unsurprised
        \item Somewhat Not Surprised
        \item Extremely Not Surprised
    \end{itemize}

    \item[Q3] Are you surprised an app can use code to make this inference?\\
    Options:
    \begin{itemize}
        \item Extremely Surprised
        \item Somewhat Surprised
        \item Neither Surprised Nor Unsurprised
        \item Somewhat Not Surprised
        \item Extremely Not Surprised
    \end{itemize}

    \item[Q4] Which of these categories describe this inference?\\
    Options:
    \begin{itemize}
        \item Identity (e.g., race, gender, age)
        \item Activities / interests
        \item Past or current location
        \item Health
        \item Relationships
        \item Education / career
        \item Finances
        \item Beliefs / politics
        \item Personally Identifiable Information (e.g., email, phone, DOB, name, SSN)
    \end{itemize}

    \item[Q5] Does this inference include other people's data besides your own?\\
    Options:
    \begin{itemize}
        \item Yes
        \item No
        \item I Don't Know
    \end{itemize}

    \item[Q6] Are you comfortable sharing this inference text with the research team?\\
    Options:
    \begin{itemize}
        \item Yes
        \item No
    \end{itemize}
\end{description}

\subsection{Pre/Post Survey Instrument}
\label{pre-post-qs}
\footnotetext{
Unless otherwise specified, agreement-based questions use a 7-point Likert scale:
\emph{Strongly disagree}, \emph{Disagree}, \emph{Somewhat disagree},
\emph{Neither agree nor disagree}, \emph{Somewhat agree}, \emph{Agree},
\emph{Strongly agree}.
}

\begin{enumerate}
    \item \textbf{How comfortable would you be with granting the following permissions to a social media application (e.g., Facebook, TikTok)?}
    \\(5-point Likert scale: \emph{Very comfortable, Somewhat comfortable, Neither comfortable nor uncomfortable, Somewhat comfortable, Very comfortable)} 
    \begin{itemize}
        \item Access to your camera roll
        \item Access to your calendar/reminders
        \item Access to your contacts
        \item Access to your location
    \end{itemize}

    \item \textbf{Did you complete the full application flow and receive a completion code?}

    \item \textbf{How comfortable would you be with granting the following permissions to a social media application (e.g., Facebook, TikTok)?}
    \\(5-point Likert scale: \emph{Very comfortable, Somewhat comfortable, Neither comfortable nor uncomfortable, Somewhat comfortable, Very comfortable)}    \begin{itemize}
        \item Access to your camera roll
        \item Access to your calendar/reminders
        \item Access to your contacts
        \item Access to your location
    \end{itemize}

    \item \textbf{Consumer online privacy is really a matter of consumers’ right to exercise control and autonomy over decisions about how their information is collected, used, and shared.}
    \hfill\emph{(7-point Likert scale)}

    \item \textbf{Consumer control of personal information lies at the heart of consumer privacy.}
    \hfill\emph{(7-point Likert scale)}

    \item \textbf{Companies seeking information online should disclose the way the data are collected, processed, and used.}
    \hfill\emph{(7-point Likert scale)}

    \item \textbf{A good consumer online privacy policy should have a clear and conspicuous disclosure.}
    \hfill\emph{(7-point Likert scale)}

    \item \textbf{It usually bothers me when online companies ask me for personal information.}
    \hfill\emph{(7-point Likert scale)}

    \item \textbf{When online companies ask me for personal information, I sometimes think twice before providing it.}
    \hfill\emph{(7-point Likert scale)}

    \item \textbf{It bothers me to give personal information to so many online companies.}
    \hfill\emph{(7-point Likert scale)}

    \item \textbf{I’m concerned that online companies are collecting too much personal information about me.}
    \hfill\emph{(7-point Likert scale)}

    \item \textbf{What is your age?}
    \begin{itemize}
        \item 18--24
        \item 25--34
        \item 35--44
        \item 45--54
        \item 55--64
        \item 65+
    \end{itemize}

    \item \textbf{How do you describe your gender identity?}
    \begin{itemize}
        \item Male
        \item Female
        \item Non-binary
        \item Prefer not to say
        \item Not listed above
    \end{itemize}

    \item \textbf{How would you describe your race? (Select all that apply)}
    \begin{itemize}
        \item $\square$ American Indian or Alaska Native
        \item $\square$ Asian
        \item $\square$ Black or African American
        \item $\square$ Native Hawaiian or Pacific Islander
        \item $\square$ White
        \item $\square$ Not listed above
        \item $\square$ Prefer not to respond
    \end{itemize}

    \item \textbf{Do you identify as Hispanic and/or Latino?}
    \begin{itemize}
        \item Yes
        \item No
        \item Prefer not to respond
    \end{itemize}

    \item \textbf{What is the highest level of education you have attained?}
    \begin{itemize}
        \item Less than high school
        \item High school graduate
        \item Some college
        \item 2-year degree
        \item 4-year degree
        \item Professional degree
        \item Doctorate
    \end{itemize}

    \item \textbf{What was your 2024 taxed income?}
    \begin{itemize}
        \item Less than \$10{,}000
        \item \$10{,}000--\$24{,}999
        \item \$25{,}000--\$49{,}999
        \item \$50{,}000--\$74{,}999
        \item \$100{,}000--\$149{,}999
        \item \$150{,}000 or greater
        \item Prefer not to respond
    \end{itemize}

    \item \textbf{Do you get the majority of your earnings from Prolific or similar platforms?}
    \begin{itemize}
        \item Yes
        \item No
        \item Prefer not to respond
    \end{itemize}
\end{enumerate}

\clearpage

\section{Statistical Evaluations }
\label{subsec:app-stats-modeling}

\subsection{Intervention-Level Ordinal Mixed-Effects Model}

\begin{table}
\centering
\caption{Cumulative link mixed-effects model predicting comfort with \textbf{calendar} permission sharing.}
\label{tab:clmm_permissions}
\begin{tabular}{lccc}
\toprule
                 Predictor & Estimate &     SE &                CI \\
\midrule
                       1|2 &   -2.410 &  0.208 &  [-2.817, -2.003] \\
                       2|3 &   -0.345 &  0.173 &  [-0.685, -0.005] \\
                       3|4 &    0.351 &  0.174 &    [0.010, 0.692] \\
                       4|5 &    2.251 &  0.206 &    [1.847, 2.654] \\
    Inference accuracy (z) &   -0.158 &  0.159 &   [-0.469, 0.153] \\
     Inference comfort (z) &    1.278 &  0.178 &    [0.930, 1.626] \\
               Time (post) &   -0.818 &  0.148 &  [-1.107, -0.529] \\
        Accuracy × Comfort &    0.090 &  0.144 &   [-0.191, 0.372] \\
           Accuracy × Time &   -0.030 &  0.144 &   [-0.312, 0.253] \\
            Comfort × Time &    0.543 &  0.159 &    [0.231, 0.855] \\
 Accuracy × Comfort × Time &   -0.092 &  0.133 &   [-0.352, 0.168] \\
\bottomrule
\end{tabular}
\end{table}

\begin{table}
\centering
\caption{Cumulative link mixed-effects model predicting comfort with \textbf{contacts} permission sharing.}
\label{tab:clmm_permissions}
\begin{tabular}{lccc}
\toprule
                 Predictor & Estimate &     SE &                CI \\
\midrule
                       1|2 &   -1.174 &  0.122 &  [-1.413, -0.936] \\
                       2|3 &    0.886 &  0.070 &    [0.750, 1.023] \\
                       3|4 &    1.471 &  0.002 &    [1.467, 1.476] \\
                       4|5 &    3.469 &  0.003 &    [3.464, 3.474] \\
    Inference accuracy (z) &   -0.260 &  0.002 &  [-0.265, -0.255] \\
     Inference comfort (z) &    1.344 &  0.003 &    [1.339, 1.349] \\
               Time (post) &   -0.515 &  0.003 &  [-0.520, -0.510] \\
        Accuracy × Comfort &    0.009 &  0.002 &    [0.004, 0.014] \\
           Accuracy × Time &   -0.072 &  0.002 &  [-0.076, -0.067] \\
            Comfort × Time &    0.456 &  0.003 &    [0.451, 0.461] \\
 Accuracy × Comfort × Time &   -0.058 &  0.002 &  [-0.063, -0.053] \\
\bottomrule
\end{tabular}
\end{table}

\clearpage
\begin{table}
\centering
\caption{Cumulative link mixed-effects model predicting comfort with \textbf{location} permission sharing.}
\label{tab:clmm_permissions}
\begin{tabular}{lccc}
\toprule
                 Predictor & Estimate &     SE &                CI \\
\midrule
                       1|2 &   -2.382 &  0.188 &  [-2.751, -2.012] \\
                       2|3 &   -0.455 &  0.155 &  [-0.758, -0.151] \\
                       3|4 &    0.402 &  0.155 &    [0.098, 0.706] \\
                       4|5 &    2.654 &  0.204 &    [2.253, 3.054] \\
    Inference accuracy (z) &    0.026 &  0.141 &   [-0.251, 0.302] \\
     Inference comfort (z) &    1.186 &  0.155 &    [0.882, 1.491] \\
               Time (post) &   -0.731 &  0.140 &  [-1.006, -0.456] \\
        Accuracy × Comfort &   -0.009 &  0.130 &   [-0.265, 0.246] \\
           Accuracy × Time &   -0.142 &  0.139 &   [-0.414, 0.130] \\
            Comfort × Time &    0.473 &  0.147 &    [0.185, 0.760] \\
 Accuracy × Comfort × Time &   -0.002 &  0.131 &   [-0.257, 0.254] \\
\bottomrule
\end{tabular}
\end{table}

\begin{table}
\centering
\caption{Cumulative link mixed-effects model predicting comfort with camera roll permission sharing}
\label{tab:clmm_permissions}
\begin{tabular}{lccc}
\toprule
                 Predictor & Estimate &     SE &                CI \\
\midrule
                       1|2 &   -2.758 &  0.212 &  [-3.173, -2.343] \\
                       2|3 &   -0.515 &  0.163 &  [-0.834, -0.195] \\
                       3|4 &   -0.061 &  0.161 &   [-0.377, 0.255] \\
                       4|5 &    2.484 &  0.204 &    [2.083, 2.884] \\
    Inference accuracy (z) &   -0.034 &  0.148 &   [-0.325, 0.256] \\
     Inference comfort (z) &    0.771 &  0.155 &    [0.467, 1.075] \\
               Time (post) &   -1.132 &  0.149 &  [-1.424, -0.841] \\
        Accuracy × Comfort &    0.142 &  0.134 &   [-0.121, 0.404] \\
           Accuracy × Time &   -0.219 &  0.140 &   [-0.494, 0.056] \\
            Comfort × Time &    0.642 &  0.149 &    [0.350, 0.934] \\
 Accuracy × Comfort × Time &   -0.037 &  0.127 &   [-0.287, 0.212] \\
\bottomrule
\end{tabular}
\end{table}

\clearpage

We note that the mean inference accuracy was 18.24, with a standard deviation of 4.19, and the mean comfort was 17.30, with a standard deviation of 4.41. 
As is standard for CLMMs, we assume proportional odds across thresholds; visual inspection of fitted probabilities suggested this assumption was reasonable.
We further note that our model does not consider importance of individual inferences, as it treats accuracies and comforts as summation scores pooled across individual inferences (e.g. two 3 comfort ratings on the Likert scale is the same as a 1 and a 5).
We made this decision as including individual inferences led to issues surrounding convergence. 
The final model converged and showed no evidence of singularity.

\section{Correlation Analysis Results}
\label{subsec:corr-analysis}

\begin{table}[h]
  \centering
  \caption{Spearman Correlation: Inference Accuracy vs.\ Reasoning Accuracy}
  \begin{tabular}{llll}
    \hline
    Coefficient ($\rho$) & Significance & Direction & Strength \\
    \hline
    0.5919 & p = 6.95e-220 (***) & Positive & Strong \\
    \hline
  \end{tabular}
\end{table}

\begin{table}[h]
  \centering
  \caption{Spearman Correlation: Inference Accuracy vs.\ Inference Comfort}
  \begin{tabular}{llll}
    \hline
    Coefficient ($\rho$) & Significance & Direction & Strength \\
    \hline
    0.2633 & p = 3.49e-38 (***) & Positive & Weak \\
    \hline
  \end{tabular}
\end{table}

\begin{table}[h]
  \centering
  \caption{Spearman Correlation: Inference Accuracy vs.\ Comfort, Stratified by Inference Type}
  \begin{tabular}{lrllll}
    \hline
    Inference Type & $n$ & Coefficient ($\rho$) & Significance & Direction & Strength \\
    \hline
    About Children & 50 & 0.2015 & p = 0.1605 (n.s.) & Positive & Weak \\
    Demographics & 208 & 0.2768 & p = 5.16e-05 (***) & Positive & Weak \\
    Education & 231 & 0.2907 & p = 7.07e-06 (***) & Positive & Weak \\
    Employment & 222 & 0.3255 & p = 7.14e-07 (***) & Positive & Moderate \\
    Financial Status & 1048 & 0.2031 & p = 3.22e-11 (***) & Positive & Weak \\
    Gambling & 2 & — & — & — & — \\
    Identifiers & 5 & -0.1481 & p = 0.8121 (n.s.) & Negative & Weak \\
    Interests & 206 & 0.2997 & p = 1.21e-05 (***) & Positive & Weak \\
    Intimate Content & 2 & — & — & — & — \\
    Legal Issues & 20 & 0.5836 & p = 0.0069 (**) & Positive & Strong \\
    Location & 1289 & 0.2704 & p = 4.85e-23 (***) & Positive & Weak \\
    Medical & 297 & 0.2453 & p = 1.91e-05 (***) & Positive & Weak \\
    Mental Health & 59 & 0.2698 & p = 0.0388 (*) & Positive & Weak \\
    Politics & 81 & 0.2502 & p = 0.0243 (*) & Positive & Weak \\
    Relationships & 313 & 0.2894 & p = 1.87e-07 (***) & Positive & Weak \\
    Religion & 101 & 0.3926 & p = 4.88e-05 (***) & Positive & Moderate \\
    Reproductive Health & 6 & 0.0000 & p = 1.0000 (n.s.) & None & Negligible \\
    Sexual Orientation & 32 & 0.1073 & p = 0.5588 (n.s.) & Positive & Weak \\
    Substance Abuse & 10 & 0.4336 & p = 0.2107 (n.s.) & Positive & Moderate \\
    Travel & 24 & 0.4966 & p = 0.0136 (*) & Positive & Moderate \\
    \hline
  \end{tabular}
\end{table}

\clearpage

\section{Inference Category Results}
\label{subsec:inf-category-results}

See Table ~\ref{tab:inf-types}.

\begin{table}[t!]
\centering
\begin{tabular}{lcc}
\toprule
Inference Type & Count & \% of Inferences \\
\midrule
    About Children & 52 & 2.2\% \\
    Demographics & 212 & 9.1\% \\
    Education & 182 & 7.8\% \\
    Employment & 224 & 9.6\% \\
    Financial Status & 405 & 17.4\% \\
    Benefits & 14 & 0.6\% \\
    Financial Identifiers & 15 & 0.6\% \\
    Purchases & 27 & 1.2\% \\
    Gambling & 3 & 0.1\% \\
    Identifiers & 11 & 0.5\% \\
    Intimate Content & 2 & 0.1\% \\
    Interests & 216 & 9.3\% \\
    Activities & 410 & 17.6\% \\
    Legal Issues & 20 & 0.9\% \\
    Location & 574 & 24.7\% \\
    Medical Conditions & 198 & 8.5\% \\
    Mental Health & 61 & 2.6\% \\
    Politics & 56 & 2.4\% \\
    Romantic, Familial, or Social Relationships & 263 & 11.3\% \\
    Religion & 104 & 4.5\% \\
    Reproductive Health & 6 & 0.3\% \\
    Sexual Orientation & 32 & 1.4\% \\
    Substance Abuse & 10 & 0.4\% \\
    Travel & 24 & 1.0\% \\
\bottomrule
\end{tabular}
\caption{Types of inferences produced by \system, by category. Percentages do not sum to 100\% because a single inference may be coded with multiple types.}
\label{tab:inf-types}
\end{table}

\clearpage 

\section{Demographics}
\label{appendix-demographics}
See Table ~\ref{table:demo}.

\begin{table}[t!]
\centering
\small 
\setlength{\tabcolsep}{3pt}
\begin{tabular}{|p{1.4cm}|p{2.0cm}|p{2.8cm}|p{2.2cm}|p{2.4cm}|p{2cm}|p{1.8cm}|}
 \hline
 \multicolumn{7}{|c|}{\textbf{Participant Demographics}} \\
 \hline
 \textbf{Age} & 
 \textbf{Gender} & 
 \textbf{Race \& Ethnicity} & 
 \textbf{Hispanic/ Latino} &
 \textbf{Education} & 
 \textbf{Income} & 
 \textbf{Privacy Concern} \\
 \hline
 
 \textbf{18-24} \newline (15.48\%) & 
 \textbf{Man} \newline (49.25\%) & 
 \textbf{White} \newline (67.96\%) & 
 \textbf{Hispanic} \newline (12.26\%) &
 \textbf{H.S. or below} \newline (0.86\%) & 
 \textbf{Below 10k} \newline (14.84\%) & 
 \textbf{IUIPC Score:} \\
 
 \textbf{25-34} \newline (18.49\%) & 
 \textbf{Woman} \newline (48.17\%) & 
 \textbf{African American} \newline (11.83\%) & 
 \textbf{Not Hispanic} \newline (87.53\%) &
 \textbf{H.S. graduate} \newline (14.41\%) & 
 \textbf{10-24k} \newline (13.76\%) & 
 \textbf{5.78} \newline (Scale 1-7) \\
 
 \textbf{35-44} \newline (18.92\%) & 
 \textbf{Non-Binary} \newline (2.58\%) & 
 \textbf{Asian} \newline (6.45\%) & 
 \textbf{Withheld} \newline (0.22\%) &
 \textbf{Some college} \newline (21.51\%) & 
 \textbf{25-49k} \newline (23.23\%) & \\
 
 \textbf{45-54} \newline (17.42\%) & 
 & 
 \textbf{Pacific Islander} \newline (0.22\%) & 
 &
 \textbf{Two year deg.} \newline (13.55\%) & 
 \textbf{50-74k} \newline (17.85\%) & 
 \\
 
 \textbf{55-64} \newline (21.08\%) & 
 & 
 \textbf{Native American} \newline (1.51\%) & 
 &
 \textbf{Four year deg.} \newline (34.19\%) & 
 \textbf{75-99k} \newline (9.03\%) & 
 \\
 
 \textbf{65-plus} \newline (8.60\%) & 
 & 
 \textbf{Two or more} \newline (8.60\%) & 
 &
 \textbf{Prof. deg.} \newline (13.76\%) & 
 \textbf{100-149k} \newline (10.54\%) & 
 \\
 
 & 
 & 
 \textbf{Other} \newline (2.80\%) & 
 &
 \textbf{Doctorate} \newline (1.72\%) & 
 \textbf{150k+} \newline (6.88\%) & 
 \\
 
 & 
 & 
 \textbf{Withheld} \newline (0.65\%) & 
 &
 & 
 \textbf{No response} \newline (3.87\%) & 
 \\
 
 \hline
\end{tabular}
\caption{\label{table:demo} Participant demographic information}
\end{table}

\clearpage
\section{Inaccuracy Types (Full)}
\label{tab:inaccuracies-appendix}
An overview of identified inaccuracy categories is provided in Table ~\ref{tab:inaccuracies-appendix}.

\vspace{.5cm}

\resizebox{1.75\linewidth}{!}{%
\begin{tabular}{|c|c|c|c|c|}
\hline
\makecell{Inacc. \\ Group} & \makecell{Inacc. \\ Type} & \makecell{\% of\\ Inacc} & Definition & Participant Example \\
\hline
                         \multirow{6}{*}{\makecell{Accurate-\\Adjacent}}  & \makecell{Wrong Person} & 11.9\% & \makecell{Correct but applied to \\ the wrong person.} & \makecell{``My husband has a moderately\\high income. I do not.} \\
                        \cline{2-5}
                         & \makecell{Wrong Relationship} & 4.5\% & \makecell{Confuses the relationship\\ between two people} & \makecell{``That isn't my message.\\It was a message sent to me.} \\
                        \cline{2-5}
                         & \makecell{Wrong Date/Time} & 18.7\% & Misinterprets timing & \makecell{``[I] was involved but no longer am.} \\
                        \cline{2-5}
                         & \makecell{Assumes Stronger} & 6.4\% & \makecell{Assumes user feels more\\strongly than they do} & \makecell{``I would not say it was a strong\\connection, just a connection.} \\
                        \cline{2-5}
                         & \makecell{Assumes Closer} & 3.5\% & \makecell{Assumes user is closer to\\someone than they are} & \makecell{``I don't have a close relationship\\with them.} \\
                        \cline{2-5}
                         & \makecell{Assumes More} & 3.0\% & \makecell{Assumes a greater quantity \\or frequency} & \makecell{``I didn't plan many birthday\\parties. Probably just one.} \\
                        \hline
                         \multirow{5}{*}{\makecell{Incorrect\\Assumption}}  & \makecell{Assumes Current\\Location Home} & 3.4\% & \makecell{Assumes user's current\\location is their home} & \makecell{``I am currently in a doctor's\\office, not my home...} \\
                        \cline{2-5}
                         & \makecell{Don't Own Home} & 4.3\% & \makecell{Assumes user owns their\\home} & \makecell{``I currently rent.\\I have never owned a home} \\
                        \cline{2-5}
                         & \makecell{Wrong Income\\Assumption} & 33.3\% & \makecell{Assumes incorrect\\financial circumstances} & \makecell{``I'd say my income is\\more middle class.} \\
                        \cline{2-5}
                         & \makecell{Wrong Demographic\\Assumption} & 8.6\% & \makecell{Assumes incorrect demographic\\(age, race, gender, etc)} & \makecell{``I am 22, so I am slightly\\below the described age bracket.''} \\
                         \cline{2-5}
                         & \makecell{Based on Defaults} & 4.9\% & \makecell{Generalization based on \\ community data} & \makecell{``although the region is Democratic,\\ (I am) not associated with this party.''} \\
                        \hline
                         \multirow{3}{*}{\makecell{Incorrect}}  & \makecell{Overly Broad} & 2.5\% & Inference is too general to be accurate & \makecell{``This is a vague generalization so\\ I cannot accurately asset it.''} \\
                        \cline{2-5}
                         & \makecell{Incorrect Interpretation} & 4.5\% & \makecell{Incorrect interpretation of\\ an acronyms, titles, etc} & \makecell{``It\'s mixing up a test \\ for a relationship.''} \\
                        \cline{2-5}
                         & \makecell{Incorrect Info Used} & 10.3\% & \makecell{Assumption based on \\ an incorrect fact} & \makecell{``The address is in the city, \\ not suburbs.''} \\
\hline
\end{tabular}
}

\clearpage

\section{Additional Plots}

\begin{figure}[!ht]
    \centering
    \includegraphics[width=.99\linewidth]{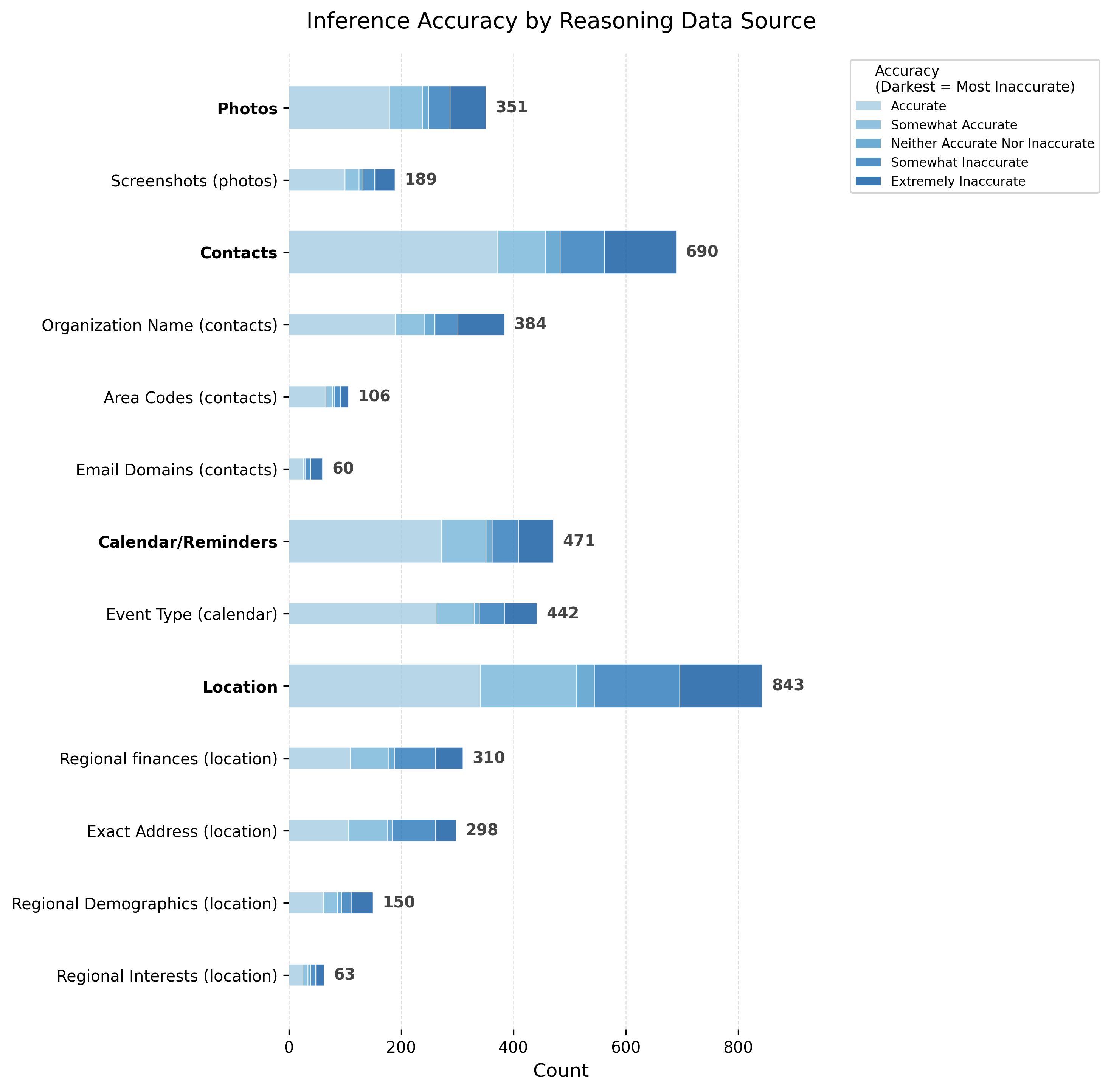}
    \caption{Inference accuracy by reasoning data source}
    \label{fig:appendix-inf-accuracy-by-reasoning-type}
\end{figure}

\begin{figure}[!ht]
    \centering
    \includegraphics[width=0.99\linewidth]{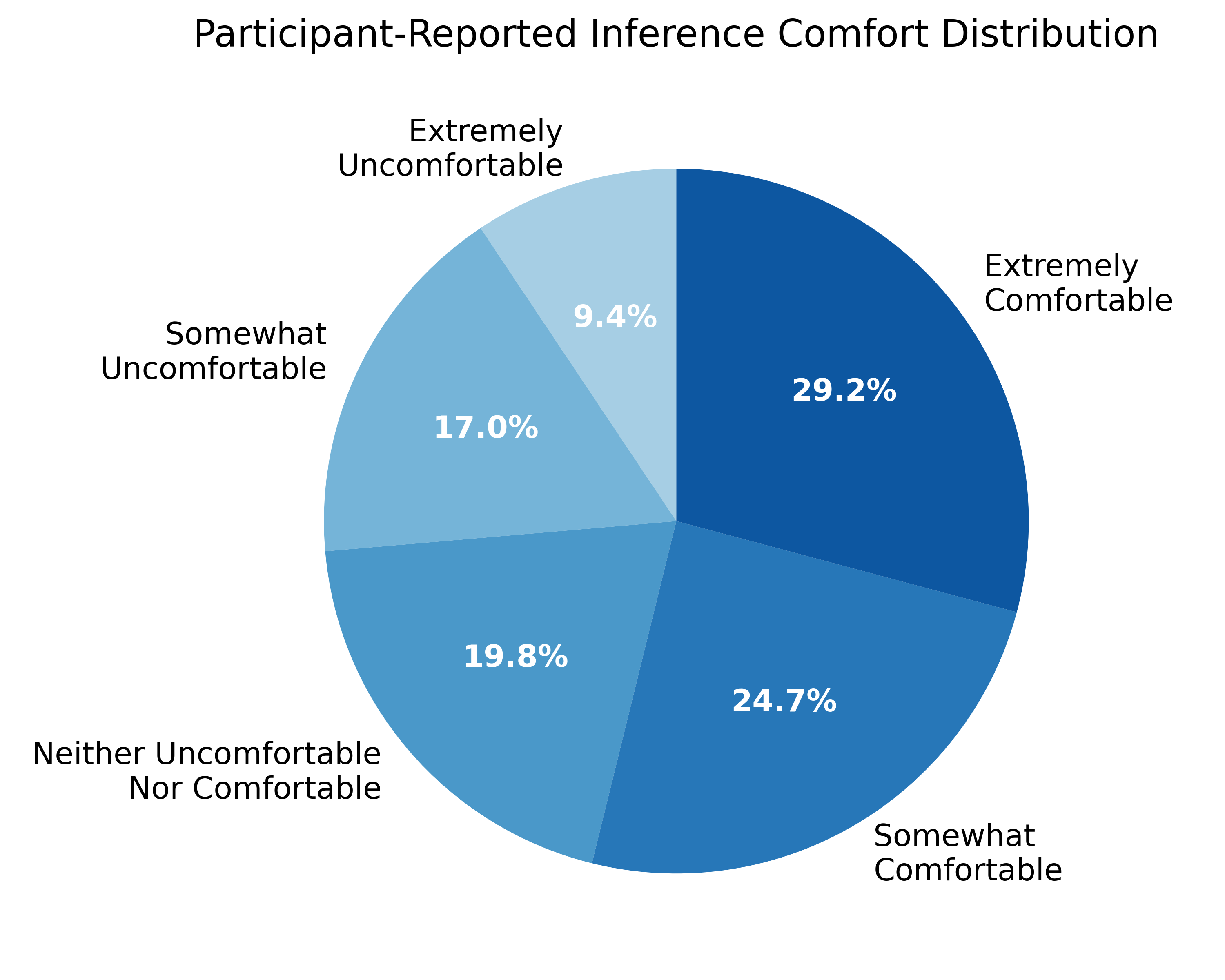}
    \caption{User comfort with inferences }
    \label{fig:comfort-pie}
\end{figure}

\begin{figure}[!ht]
    \centering
    \includegraphics[width=0.99\linewidth]{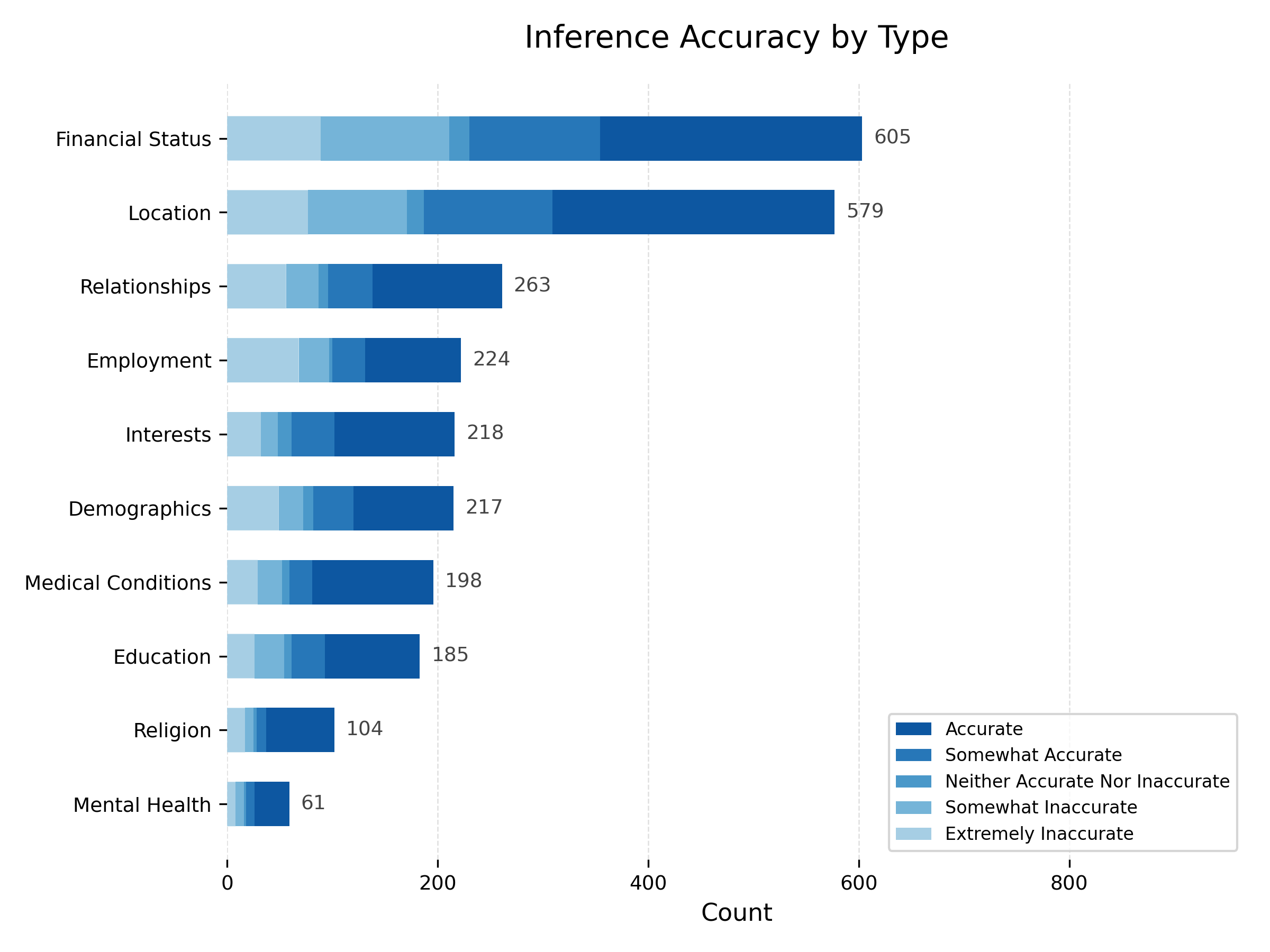}
    \caption{User comfort levels by inference type, when asked ``Are you comfortable with this inference made by AI?'' }
    \label{fig:inf-comfort}
\end{figure}

\begin{figure}[!ht]
    \centering
    \includegraphics[width=0.99\linewidth]{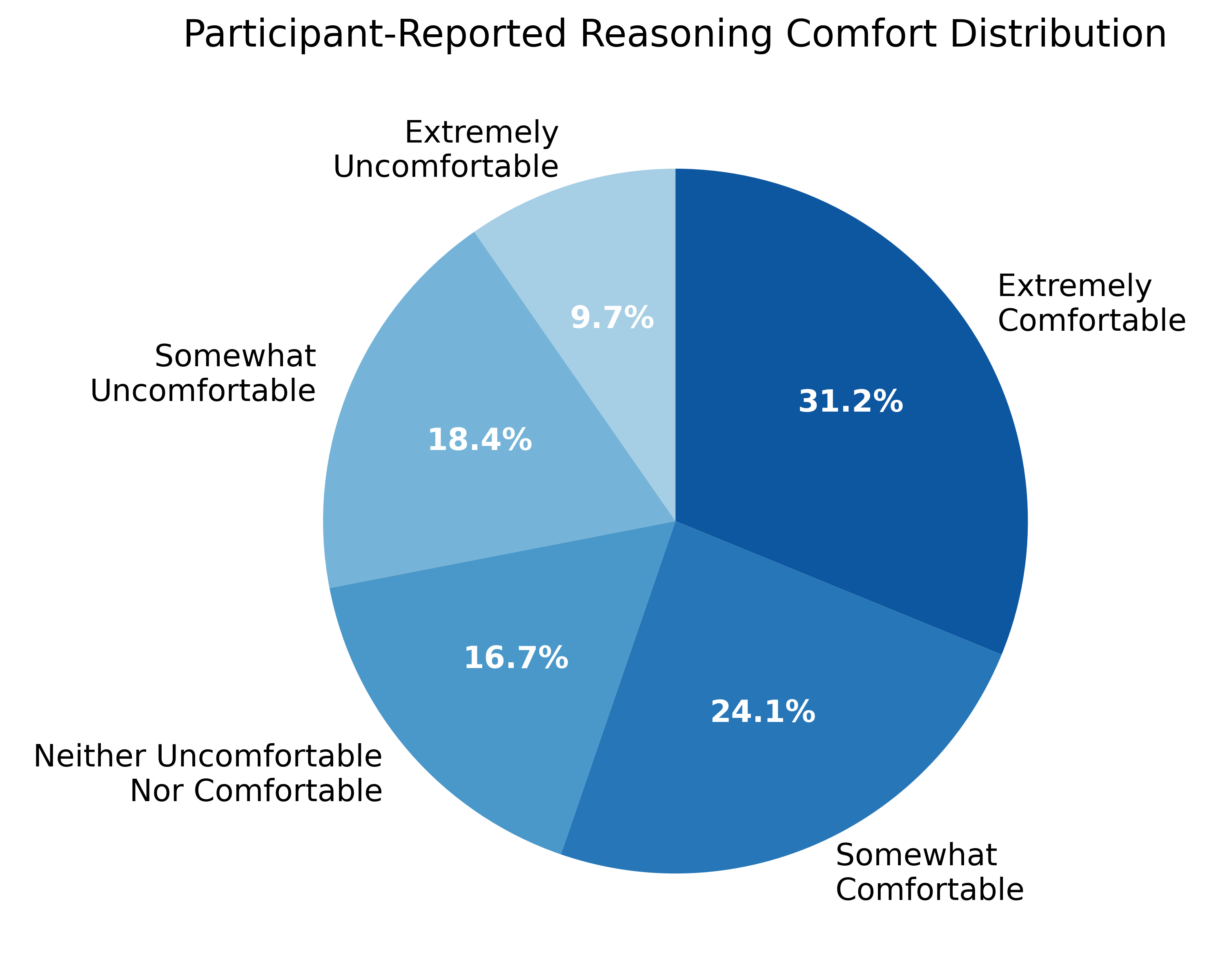}
    \caption{User comfort with reasoning}
    \label{fig:reas-comfort-pie}
\end{figure}